\documentclass[useAMS,usenatbib]{mn2e}
\usepackage{graphicx}
\usepackage{amsmath}
\usepackage{txfonts}

\def\lya{Ly$\alpha$}

\title[Ly$\alpha$ LFs  at  $z\approx 4.5$]{Ly$\alpha$ Luminosity Functions at Redshift $z\approx 4.5$}

\author[Z.Y. Zheng et al. 2012]
{
Zhen-Ya Zheng$^{1,2}$\thanks{E-mail:zhenya.zheng@asu.edu}, 
Steven L. Finkelstein$^{3,7}$\thanks{Hubble Fellow}, 
Keely Finkelstein$^{3}$,  
Vithal Tilvi$^{4}$, 
\newauthor
James E. Rhoads$^{2}$, 
Sangeeta Malhotra$^{2}$,
Jun-Xian Wang$^{1}$,
Neal Miller$^{5}$,
\newauthor
Pascale Hibon$^{6}$, and
Lifang Xia$^{2}$.\\
$^{1}$CAS Key laboratory for Research in Galaxies and Cosmology, Department of Astronomy,
 University of Science and Technology of China, Hefei, Anhui 230026, China\\
$^{2}$School of Earth and Space Exploration, Arizona State University, Tempe, AZ 85287\\
$^{3}$Department of Astronomy, The University of Texas at Austin, Austin, TX 78712\\
$^{4}$Department of Physics, Texas A\&M University, College Station, TX 77843\\
$^{5}$Department of Astronomy, University of Maryland, College Park, MD 20742\\
$^{6}$Gemini Observatory, La Serena, Chile 
}

\begin{document}

\date{Accepted XXXX. Received XXXX; in original form XXXX}

\pagerange{\pageref{firstpage}--\pageref{lastpage}} \pubyear{xxxx}

\maketitle

\label{firstpage}

\begin{abstract}
We present a spectroscopically confirmed sample of Lyman alpha
emitting galaxies (LAEs) at $z \sim 4.5$ in the Extended Chandra
Deep Field South (ECDFS), which we combine with a sample of 
$z\sim 4.5$ LAEs from previous narrowband surveys
 from the Large Area Lyman Alpha (LALA) survey to build a
unified \lya\ luminosity function.  We spectroscopically observed 64
candidate LAEs in the ECDFS, confirming 46 objects as z $\sim$ 4.5
LAEs based on single-line detections with no continuum
emission blue-ward of the line, resulting in a \lya\ confirmation
rate of $\sim$ 70\%. We did not detect significant flux from neither the
C\,{\sc iv} $\lambda$1549\AA\ emission line nor the He\,{\sc ii} $\lambda$1640\AA\ emission
line in individual LAE spectra.  These lines were also undetected in a coadded spectrum, with the
coadded line ratio of He\,{\sc ii} to \lya\ constraining the Population III star
formation rate to be $<$0.3\% of the total star formation rate (SFR), and $<$1.25\% of the observed SFR (both at the 2-$\sigma$ level). We combine
the optical spectra with deep X-ray and radio images to constrain
the AGN fraction in the sample.  Only LAE was detected in both the X-ray and radio, 
while the other objects remained undetected, even when stacked.
The \lya\ luminosity functions in our two deepest narrowband filters in the ECDFS differ
at greater than 2$\sigma$ significance, and the product $L^*\Phi^*$ differs by a factor of $>3$. 
Similar luminosity function differences have been used to infer evolution in the neutral gas fraction in the intergalactic medium at $z>6$, 
yet here the difference is likely due to cosmic variance, given that the two samples are from
adjoining line-of-sight volumes.
Combining our new sample of LAEs with those from previous LALA narrowband surveys at z = 4.5, we obtain one of
the best measured \lya\ luminosity functions to date, with our sample of over 200
spectroscopically confirmed \lya\ galaxies yielding
log$_{10}$(L$^{*}$) = 42.83 $\pm$ 0.06 [ergs $s^{-1}$] and
log$_{10}$($\Phi^*$) = -3.48 $\pm$ 0.09 [Mpc$^{-3}$].
We compare our new luminosity function to others from the literature to study the
evolution of the \lya\ luminosity density at $0 < z < 7$.
We find tentative evidence for evolution in the product $L^* \Phi^*$, which approximately
tracks the cosmic star formation rate density, but since
field-to-field and survey-to-survey variations are in some cases as
large as the possible evolution, some caution is needed in interpreting
this trend.
\end{abstract}


\section{INTRODUCTION}
\label{sec:intro}
Thanks to efficient wide field cameras, the number of star-forming galaxies known in the early universe has grown rapidly in the past decade. These high-redshift star-forming galaxies are selected mainly through two techniques, the dropout technique and the \lya-line search technique. The former is known as the Lyman-break technique (Steidel et al. 1996), and is applied using deep broadband images wherein high-redshift galaxies are identified via a flux discontinuity caused by absorption from neutral gas in the inter-galactic medium (IGM).   The latter method is designed to search for the strong \lya\ emission line, using deep narrowband images to identify galaxies where the \lya\ line is redshifted to windows of low night-sky emission. These are known as \lya\ emitting galaxies (LAEs) and have been found at multiple redshifts from z = 2.1 to z =7.3 (e.g., Cowie \& Hu 1998, Guaita et al 2010, Gawiser et al. 2007, Rhoads et al. 2000, 2003, Ouchi et al. 2008, Wang et al. 2005, Dawson et al. 2004,2007, Iye et al. 2006, Finkelstein et al. 2008, 2009b,c,  Hibon et al 2011, Shibuya et al. 2012, and Rhoads et al. 2012). There have also surveys which have identified candidate  LAEs at z $\geq$ 7.7 (e.g., Tilvi et al 2010, Krug et al 2012), but no  spectroscopic confirmations have yet been reported for LAEs at z $>$ 7.3. 
LAE searches by definition identify galaxies with strong \lya\ line emission, but they also allow the discovery of galaxies with rest-frame ultraviolet (UV) continuum levels which may be too faint to be detected via the Lyman-break technique, allowing one to study the faint-end of the observable galaxy population.  High-redshift LAEs identified via narrowband surveys typically have a spectroscopic confirmation success fraction \footnote{The success fraction is defined as the ratio of number of spectroscopically confirmed LAEs to the number of effective targets.  The effective targets exclude those where the expected line location was contaminated by strong residuals of night sky emission lines, or fell in a chip gap in the CCD mosaic.} of $\gtrsim$ 70\% (Dawson et al. 2007; Wang et al. 2009).  Although narrowband surveys have found thousands of \lya\ emitters from $z =$ 2.1--6.96, fewer than 1000 LAEs have been spectroscopically confirmed. 

While the rest-frame UV continuum luminosity function of high-redshift galaxies changes markedly from 2 $< z <$ 7 (e.g., Bouwens et al.\ 2007), the luminosity function of the \lya\ emission line of LAEs is essentially non-evolving over the redshift range z $\sim$ 3 to z $\sim$ 5.7 (e.g., Malhotra et al. 2012, Malhotra \& Rhoads 2004, Dawson et al. 2007, Wang et al. 2009, Ouchi et al. 2008), where their luminosity density peaks.  They show a slight ($\sim 30\%$) decrease from z $\sim$ 5.7 to z $\sim$ 6.5 \citep{Ouchi10, Kashikawa11, Hu10}, and they show a possible decrease from $z \sim 3$ to $z \sim 2$ \citep{Ciardullo12}, though with current sample sizes these changes remain at the $\sim 3\sigma$ level. The lower-redshift evolution might be due to the rise of active galactic nuclei (AGNs) in the LAE samples at redshift z = 2.1 (e.g., Nilsson \& M{\o}ller 2011), but these recent results highlight the need to perform a more in-depth examination of the evolution of the \lya\ luminosity functions for LAEs between redshift 3.1 and 5.7.

Any AGNs present in LAE samples can be detected through the presence of high ionization lines, e.g., C\,{\sc iv} in the rest-frame UV. Other lines such as He,{\sc ii} may also indicate AGN activity, though this line is also theorized to be a byproduct of the hard continuum expected from the first generation of metal-free stars (Schaerer 2002).
No C\,{\sc iv}  or He\,{\sc ii} emission lines were reported in previous LAE studies, even in coadded spectra \citep{Nagao05,Dawson07,Ouchi08,Wang09}.  Deep X-ray images are the most effective way to find AGNs, though the large distance to these LAEs renders X-ray studies able to find only luminous AGN.
  
  Previous 170ks and 180ks {\it Chandra} X-ray exposures in the Large Area Lyman Alpha \citep[LALA,][]{Rhoads00} fields did not find any X-ray individual or average detections in LAEs at z$\sim$4.5 \citep{Malhotra03,Wang04}. However, in the ECDFS region, which contains 250ks X-ray imaging over the full field, as well as the deepest X-ray imaging in the sky of 4Ms in the Chandra Deep Field South (CDF-S), we previously found one quasar in out of 112 LAE candidates at z$\sim$4.5 \citep{Zheng10}. Excluding that LAE-AGN, the remaining LAEs did not show any detection even when stacked. At redshift 2 $\leq$ z $\leq$4, more AGNs are found through deep X-ray surveys, but the AGN fraction is still low \citep[$\leq$5\%,][]{Gawiser07, Guaita10, Ouchi08}. This differs from the z $\sim$ 0.3 LAEs selected via GALEX spectroscopy, where the AGN fraction is as high as 15\%-40\%, though more than just X-ray imaging is used to identify possible AGNs \citep{Finkelstein09a,Scarlata09,Cowie10}.

Here we present new spectroscopic observations of z $\sim$ 4.5 LAEs in the 0.34 deg$^2$ ECDFS region, and combine this new sample with previous LALA Bootes \citep{Dawson04,Dawson07} and LALA Cetus \citep{Wang09} LAE samples to derive a unified \lya\ luminosity function at z$\sim$4.5.  This region has an extraordinary amount of multi-wavelength data, including high-resolution {\it Hubble Space Telescope} optical and {\it Spitzer Space Telescope} infrared images from the Great Observatories Origins Deep Survey \citep[GOODS;][]{Giavalisco04}, deep ground-based $UBVRIzJHK$ photometry from the Multiwavlength Survey by Yale-Chile (MUSYC) \citep{Gawiser06} and the ESO Imaging Survey (EIS) \citep{Arnouts01}, deep radio data from the Very Large Array (VLA; Miller et al. 2008, Miller et al. in Prep.), and deep X-ray data from {\it Chandra} (Giacconi et al. 2002; Alexander et al. 2003; Lehmer et al. 2005; Luo et al. 2008; Xue et al. 2011). Our candidates and spectroscopic targets are primarily selected from two narrowband images centered at wavelengths to select z $\sim$ 4.5 LAEs (first presented in Finkelstein et al. 2009a), combined with the EIS R-band data. We present our photometric and spectroscopic observations in \S 2, spectroscopic results in \S 3, and discuss the AGN contamination fraction, the constraints on Population III stars, and the properties of \lya\ luminosity functions in \S 4. Throughout this work, we assume a cosmology  with $H_0$ = 70 km s$^{-1}$ Mpc$^{-1}$, $\Omega_m$ = 0.27, and $\Omega_\Lambda$ = 0.73 \citep[c.f.][]{Komatsu11}. At redshift $z$ = 4.5, the corresponding age of the universe was 1.36 Gyr old, with a scale of 6.7 kpc/\arcsec, and a redshift change of $\delta$z = 0.03 implies a comoving distance change of 18.75 Mpc \citep[based on Ned Wright's cosmology calculator,][]{Wright06}. Magnitudes are given in the $AB$ system.

\section{OBSERVATIONS}
\label{sec:obs}

\subsection{Candidate Selection through Narrowband Imaging}
\label{sec:obs:sel}

The LAE candidates were selected using narrowband imaging of the GOODS {\emph{Chandra Deep Field} South (CDF-S; RA 03:31:54.02, Dec $-$27:48:31.5, J2000)  obtained at the Blanco 4m telescope at the Cerro Tololo InterAmerican Observatory (CTIO) with the MOSAIC II camera. Deep images were obtained in three 80 \AA\ wide narrowband filters: NB656 (2.75 hr), NB665 (5 hr) and NB673 (5.3 hr; where the filter name denotes the central wavelength in nm; Finkelstein et al. 2009c). We used the broad band R image from the ESO Imaging Survey \citep[EIS, area = 34 x 33 arcmin$^2$,][]{Arnouts01}. The EIS-R filter  has a central wavelength of 6517 \AA\ and full-width at half-maximum (FWHM) of 1622 \AA.  As the central wavelength of the R-band image is similar to our narrowband images, we used the R-band data (which has a 5$\sigma$ limit of m(R)$_{lim}$ = 25.6 measured in a 2\arcsec aperture) to measure the zero-point in our narrowband images.  The LAE candidates were selected in Finkelstein et al. (2009c), based on a 5 $\sigma$ significance detection in the narrowband (f$_{NB, \lambda}$ $>$ 5$\sigma_{NB}$), a 4 $\sigma$ significance narrowband flux density excess over the R band flux density (f$_{NB, \lambda}$ - f$_{R, \lambda}$ $>$ 4$\times$ $\sqrt{\sigma_{NB}^2+ \sigma_{R}^2}$), a factor of 2 ratio of narrowband flux density to broadband flux density (f$_{NB, \lambda}$/f$_{R, \lambda}$ $>$ 2), and no more than a 2
$\sigma$ detection in the B-band (f$_{B, \lambda}$ $<$ 2$\sigma_{B}$; using the B-band image from EIS, which has a 2-$\sigma$ magnitude limit of B$_{AB}$ = 27.4). The first three criteria ensure a significant line detection, while the last criterion is a necessary condition for objects at z $>$ 4 via the Lyman break. The 5$\sigma$ magnitude limits of the narrowband images (NB665 and NB673, with 2\arcsec aperture) of m([NB665, NB673])$_{lim}$ = [25.0, 25.2] correspond to pure emission line fluxes $>$ [2.2, 1.8] $\times$ 10$^{-17}$ ergs cm$^{-2}$ s$^{-1}$, and the factor of 2 ratio of narrowband flux density to R band flux density corresponds to EW$_{rest}(Ly\alpha)$ $\geq$ 16.2\AA. 
 With these criteria, we have selected  4 candidate LAEs in NB656 (Finkelstein et al. 2008)\footnote{The NB656 candidates were selected only in the overlap area between the shallow narrowband image and the GOODS {\it Hubble Space Telescope} data (160 arcmin$^2$), which is why only four objects were selected. The other two catalogs consist of all selected candidates over the overlap region between the deeper narrowband image and the ESO Imaging Survey, which consists of a much larger area. },  42 in NB665, and 85 in NB673. 11 candidate LAEs were selected in both NB665 and NB673, which have significantly overlapping filters (Finkelstein et al. 2009c). 
 Candidates with GOODS B-band coverage were further examined in that deeper image (B$_{AB}$ = 30.1 at the 2-$\sigma$ level, Giavalisco et al. 2004),
  and though none were formally detected at $>$2$\sigma$, 3 candidates in NB665 and 5 candidates in NB673 had visible flux in the GOODS B-band images, and thus were excluded. Once all of these criteria were applied, we were left with 2 objects from NB665 and 8 from NB673 in the GOODS area (Finkelstein et al. 2009a). In total our sample comprises 112 LAE candidates (4 in NB656, 33 in NB665, and 75 in NB673) at z $\sim$ 4.5. 

The estimated number of interlopers (e.g., [O\,{\sc ii}] and [O\,{\sc iii}] emitters) should be low, as we require a significant emission line detection in the narrowband as well as no significant detection in the B-band (H$\alpha$ emitters are also a possibility, but are unlikely since the volume at z$\sim$0 is very small). With the complete sample of emission-line galaxies from the Hubble Space Telescope Probing Evolution and Reionization Spectroscopically Grism Survey (PEARS, Pirzkal et al. 2012), the fraction of [O\,{\sc ii}] and [O\,{\sc iii}] emitters with B $\geq$ 27.4 (M$_{4350\AA}$ $>$ -17.8 for [O\,{\sc ii}] emitters, and M$_{4350\AA}$ $>$ -16.4 for [O\,{\sc iii}] emitters) are $\leq$ 3\% and $\leq$ 4\%, respectively. PEARS had found 269 [O\,{\sc ii}] emitters in the redshift range of 0.5--1.6, and 464 [O\,{\sc iii}] emitters in the redshift range of 0.1--0.9 in an area of 119.08 arcmin$^2$. Assuming that the number density of emitters did not evolve with redshift in the corresponding range, we estimate that there are $\sim$2 [O\,{\sc ii}] emitters and 3 [O\,{\sc iii}] emitters in our narrowband sample. Within the GOODS CDF-S area, the estimated number of [O\,{\sc ii}] emitters and [O\,{\sc iii}] emitters are 0.33 and 0.5.  

\renewcommand{\thefootnote}{\alph{footnote}}

\begin{table*}
\centering
\begin{minipage}{100mm}
\caption{Number of Candidate, Targeted and Confirmed \lya\ Emitters at z $\sim$ 4.5 in ECDFS field. }
\begin{tabular}{@{}lccccc@{}}
 \hline\hline
Filter & Seeing &  mag$_{limit}^a$ & Candidates$^b$ &  Targeted$^c$  & Confirmed$^c$ \\
\hline 
NB656   &		0\arcsec.90		&		 24.7		&    4	 &	 3	&	 3	\\
NB665  &		0\arcsec.92		      &		 25.0		&		 33	&	 17	&	 11	\\
NB673  &		0\arcsec.91 	&		25.2		&		 75	&	44	&	 32	\\
\hline
\label{nb}
\end{tabular}
\footnotetext[1]{ 5-$\sigma$ limiting magnitude (aperture magnitude within 2\"\ diameter) at the corresponding narrowband.}
\footnotetext[2]{ 11 LAE candidates were covered by both the NB665 and NB673 filters, and we divided them into 
the corresponding narrowband (5 into NB665 and 6 into NB673) due to their narrowband flux.} 
\footnotetext[3]{6 of 11  LAE candidates covered by both the NB665 and NB673 filters were targeted,  3 were confirmed 
as NB665 LAEs and 3 were confirmed as NB673 LAEs, and the confirmation is consistent with the narrowband division.} 
\end{minipage}
\end{table*}

\begin{table*}
\centering
\begin{minipage}{100mm}
\caption{Number of Targeted and Confirmed \lya\ Emitters at z $\sim$ 4.5 in Each Mask in the CDF-S field.  }
\begin{tabular}{@{}lcccc@{}}
 \hline\hline
Mask 	&	Grism (line/mm)	&	 Exp. Time (ks)	&	 Targeted 	&	Confirmed \\
\hline 
 1 		&	200		&	 10.2		&	20 		&	11	\\ 
2 		&	200 		&	 12.2		&	26 		&	20 	\\
3		&	 300 	&	 16.2		&	 22 		&	15 	\\
4 		&	300 		&	 14.4		&	16		&	15 	\\
5		 &	300 		&	 14.4		&	10 		&	  7	\\ \hline
\label{specobs}
\end{tabular}
\end{minipage}
\end{table*}

\subsection{Spectroscopic Observations}
\label{sec:obs:spec}

Our spectroscopic data were obtained using the IMACS (Dressler et al. 2006) short camera ($f/2$, with a 27\arcmin .2 diameter field of view)  on the 6.5 m Magellan I Baade Telescope in 2009 September 10-11 (through Steward Observatory time, PI Rhoads) with the 200 line/mm grism  and 2009 November 11-12 (NOAO PID 2009B-0371, PI Finkelstein) with the 300 line/mm grism.  The 200 line/mm grism has $\lambda_{blaze}$ = 6600 \AA\ and a resolution of 2.037 \AA\ pixel$^{-1}$ 
 with a range of 4000-10500 \AA, and the 300 line/mm  grism has $\lambda_{blaze}$ = 6700 \AA, and a resolution of 1.341 \AA\ pixel$^{-1}$ with a range of 4000-9200 \AA.  Five multi-slit masks (see Table 2) were observed for 10.2-16.2 ks with 0.5 hr individual integration times. The masks have slit widths of 0.8 arcsec. Each slit mask included approximately 20  candidate \lya\ emitters (mixed in with roughly $\sim$150 [with 200 line/mm grism] or $\sim$50 [with 300 line/mm grism] other spectroscopic targets).  Of these, 16 candidates were covered by more than one mask. In total we targeted 64 LAE candidates.

The data were reduced using the IMACS version of the Carnegie Observatories System for MultiObject Spectroscopy (COSMOS) data reduction package\footnote{http://obs.carnegiescience.edu/Code/cosmos/Cookbook.html}. We first determined two-dimensional wavelength solutions for each science exposure using arc lamp exposures taken immediately before or after each science frame. The wavelength residuals in the calibration is $\sim$2 pixels. After wavelength calibration, each frame was first bias-subtracted and flat-fielded. We then performed sky subtraction following the algorithm described by Kelson (2003), which modeled the camera distortions and the curvature of the spectral features in the two-dimensional background spectrum for subtraction, and extracted one-dimensional spectra from the two-dimensional spectra using the task "extract-2dspec" for each slit. 

To control for possible spatial shifts along the slits between individual exposures of a mask, we measured the trace locations of the brightest continuum sources in the mask. We corrected for any measured shifts while stacking the exposures for each mask to increase the quality of the stacked two-dimensional spectra. We also identified and removed cosmic ray hits while stacking the multiple exposures of each mask. 

\begin{figure*}
\begin{minipage}{180mm}
\includegraphics[totalheight=0.8\textheight]{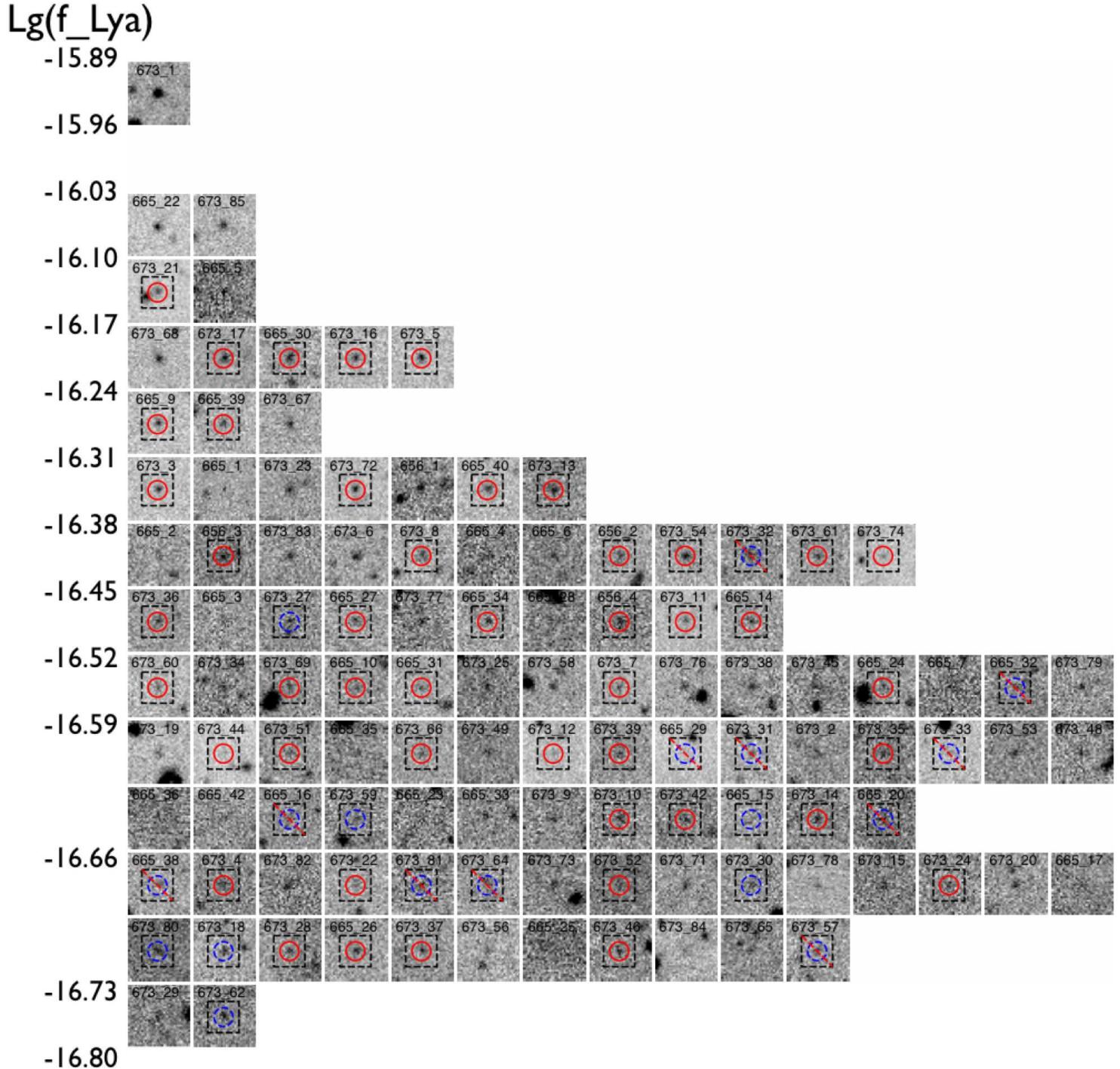}
\caption{ Narrowband stamps (size: 10\arcsec x 10\arcsec, 0.27\arcsec per pixel) of  all 112 LAE candidates, 
including 46 spectroscopically confirmed LAEs with red circles, 18 spectroscopically observed yet un-confirmed candidates with blue circles, and 48 non-targeted 
 candidates without any marks. In the sample of not confirmed \lya\ emitter candidates, 11 objects marked as blue circles with red-dashed diagonal 
lines show no emission lines in their spectra, 7 objects marked as blue circles without red-dashed diagonal 
lines show continuum with little-to-no \lya\ emission line (except 673-27 which exhibits a strange profile and is hard to judge). 
All stamps are divided into 13 rows (row 2 is empty, and rows 11 and 12 have 27 and 26 candidates, respectively), with the rows representing a decreasing \lya\ line flux of 0.07 dex. Note that as expected, the candidates with larger $f_{Ly\alpha}$ are confirmed at a higher rate.}
\label{stamp}
\end{minipage}
\end{figure*}

\begin{figure}
\begin{center}
\includegraphics[totalheight=0.3\textheight]{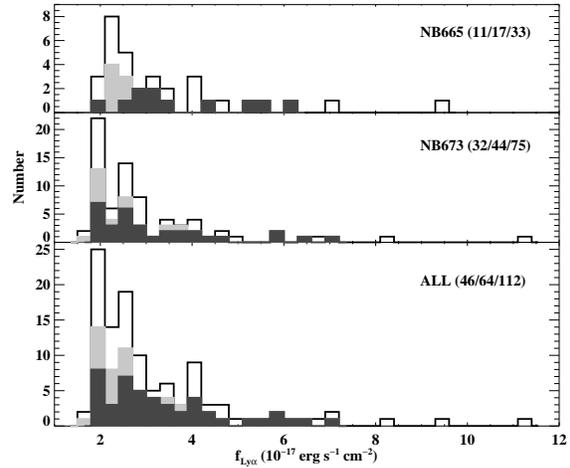} 
 \caption{\lya\ flux (computed from the narrowband excess) distributions of all candidate LAEs (empty histogram),
 IMACS targeted LAEs (grey filled histogram) and spectroscopically confirmed LAEs
(blacked filled histogram). Candidates selected in the two narrowband images (NB665 and NB673) are plotted in the middle and 
top panels. The 11 LAE candidates covered in both NB665 and NB673 are divided into the corresponding narrowband based on their \lya\ fluxes in the two narrowband filters. } 
\label{nbdist}
\end{center}
\end{figure}

\begin{figure}
\begin{center}
\includegraphics[totalheight=0.3\textheight]{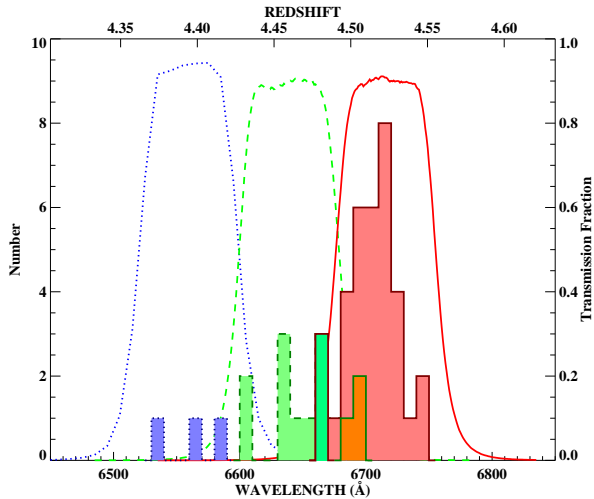} 
\caption{Narrowband filter transmission curves (bottom/right axes) and redshift distributions of confirmed LAEs (top/left axes). The histograms denote the redshift distributions of confirmed LAEs, where the colors denote the narrowband detection filter (NB656, NB665, and NB673 from left to right), using the same colors as the filters' transmission curves,  Orange and light green filed histograms are confirmed candidates observed in both the NB673 and NB665 filters. The NB656 candidate LAEs are selected with a much smaller area and shallower narrowband image compared to the NB665 and NB673 LAEs. } 
\label{filterz}
\end{center}
\end{figure}

\section{SPECTROSCOPIC RESULTS} 
\label{sec:spec}

\subsection{Line Identification}
\label{sec:spec:line}
Among our 64 candidate LAEs observed, 46 objects exhibiting a single emission line consistent with \lya\ at $z \sim$ 4.5 (see figure \ref{conf1} and \ref{conf2} about their 1-d and 2-d spectra).  Out of the remaining 18 candidates, 1 is likely lost in the CCD gaps (NB665-15), 2 are contaminated by sky emission lines (NB665-32 and NB665-38) from nearby slits, 9 show no detection of continuum or emission lines, 5 show  continuum breaks with little-to-no \lya\ line, and one object (NB673-27) may be a LAE, though if so it has a strange line profile.  We thus find a spectroscopic confirmation success rate of $\sim$70\%-80\% (the exact rate is 71.9\%; the range is estimated by including or excluding the first four LAEs of the 18 unconfirmed candidates), and the contamination fraction is about 14\%-19\%. 
We present a catalog of the 46 confirmed z $\sim$ 4.5 LAEs in Table \ref{data}, and their redshift distribution is shown in figure \ref{filterz}. Six out of the 11 LAE candidates detected in both the NB665 and NB673 images were targeted with IMACS.  Based on the spectroscopic redshift, three of these were confirmed as NB665 LAEs and three were confirmed as NB673 LAEs, consistent with their narrowband fluxes in the respective filters (see figure \ref{filterz}). In Figure \ref{stamp}, we present narrowband images of all candidate LAEs (labeling those which were targeted spectroscopically, as well as those that were confirmed to be LAEs). We show the distribution of \lya\ fluxes in Figure \ref{nbdist}. All targeted LAE candidates with f$_{Ly\alpha}$ $>$ 3.7$\times10^{-17}$ erg cm$^2$s$^{-1}$ are confirmed as LAEs.

Out of the 8 candidates which appear to show flux in the GOODS B-band image (though do not exhibit a formal detection) , we targeted three of them, and found one to be an [O\,{\sc iii}] emitter with very narrow (observed frame width $<$ 6\AA) [O\,{\sc iii}] $\lambda$5007 \AA\ , marginal [O\,{\sc iii}] $\lambda$4959\AA\ and marginal H$\beta$ $\lambda$4861\AA\ emission lines. This number is consistent with our estimation of the interloper fraction. The other two candidates showed continuum in their 2D spectra.

22 candidate LAEs were observed more than once. Four of them show no line but do show continuum flux and are excluded as LAEs. Two LAEs observed twice were confirmed with only a single observation, since their other spectrum was affected by the CCD gap or contamination from a nearby bright star.  The redshift values estimated from the line peak of the remaining 16 multi-frame confirmed LAEs show consistency within 1 pixel.  The four candidates that show continuum but no line are all visible in the grism-300 2D spectra, while only two can be resolved in the grism-200 spectra. From table 2 we can see the first day observation with grism-200 shows a relatively low success-fraction. This is likely due to the lower exposure time of this observation (2ks less than the other mask observations).

Although the objects identified via a single emission line are likely LAEs, so to increase our confidence in this result, we examine the shape of the detected emission line.  We typically identify \lya\ emission by its characteristically asymmetric line profile, as the blue wing of the \lya\ line can be absorbed by neutral hydrogen in the IGM. Following Rhoads et al. (2003, 2004), we perform two measurements of the line asymmetry: the wavelength-based asymmetry defined as $a_\lambda$ = $(\lambda_{10,r}-\lambda_p)/(\lambda_p-\lambda_{10,b})$, and the flux-based asymmetry defined as $a_f$ = ($\int_{\lambda_p}^{\lambda_{10,r}}f_\lambda d\lambda )/ (\int_{\lambda_{10,b}}^{\lambda_p}f_\lambda d\lambda $), where the $\lambda_p$ is the wavelength of the emission-line peak, and the $\lambda_{10,b}$ and $\lambda_{10,r}$ are the wavelengths where the flux density equals 10\% of the peak on the blue side and red side, respectively. In Figure \ref{asym}, we plot $a_\lambda$ versus $a_f$ for the 46 spectroscopically confirmed LAEs. The error bars on $a_\lambda$ and $a_f$ are estimated based on 1000  Monte-Carlo simulations, in which we added random noise (proportional to the flux error) to each data bin, and re-measured the asymmetry parameters. There are 34 LAEs which show both $a_\lambda$ and  $a_f$ greater than 1, and only 2 LAEs show both $a_\lambda<1$ and $a_f < 1 $ at $>$ 1$\sigma$ significance. These two LAEs (CH8-1/NB665-24 and NB673-72) have symmetric 2D spectra. We continue to include these objects in our analysis; an unusual line profile is not by itself enough to rule out a \lya\ line identification given the range of possible line profiles produced by \lya\ radiative transfer (e.g., Zheng, Cen, et al. 2010). We also notice that the asymmetry measured with the 300 line/mm grism shows less dispersion than that measured with the 200 line/mm grism.  This is likely due to longer exposure times and better spectral resolution.

We also search for other emission lines (e.g., C\,{\sc iv}$\lambda$1549\AA, and He\,{\sc ii}$\lambda$1640\AA) in the spectra of our confirmed LAEs. However, we do not find any detections, though these lines are likely too faint to detect in our spectra due to the limited depth, which was designed to detect the strong \lya\ emission line.

\begin{figure*}
\begin{center}
\includegraphics[totalheight=.9\textheight]{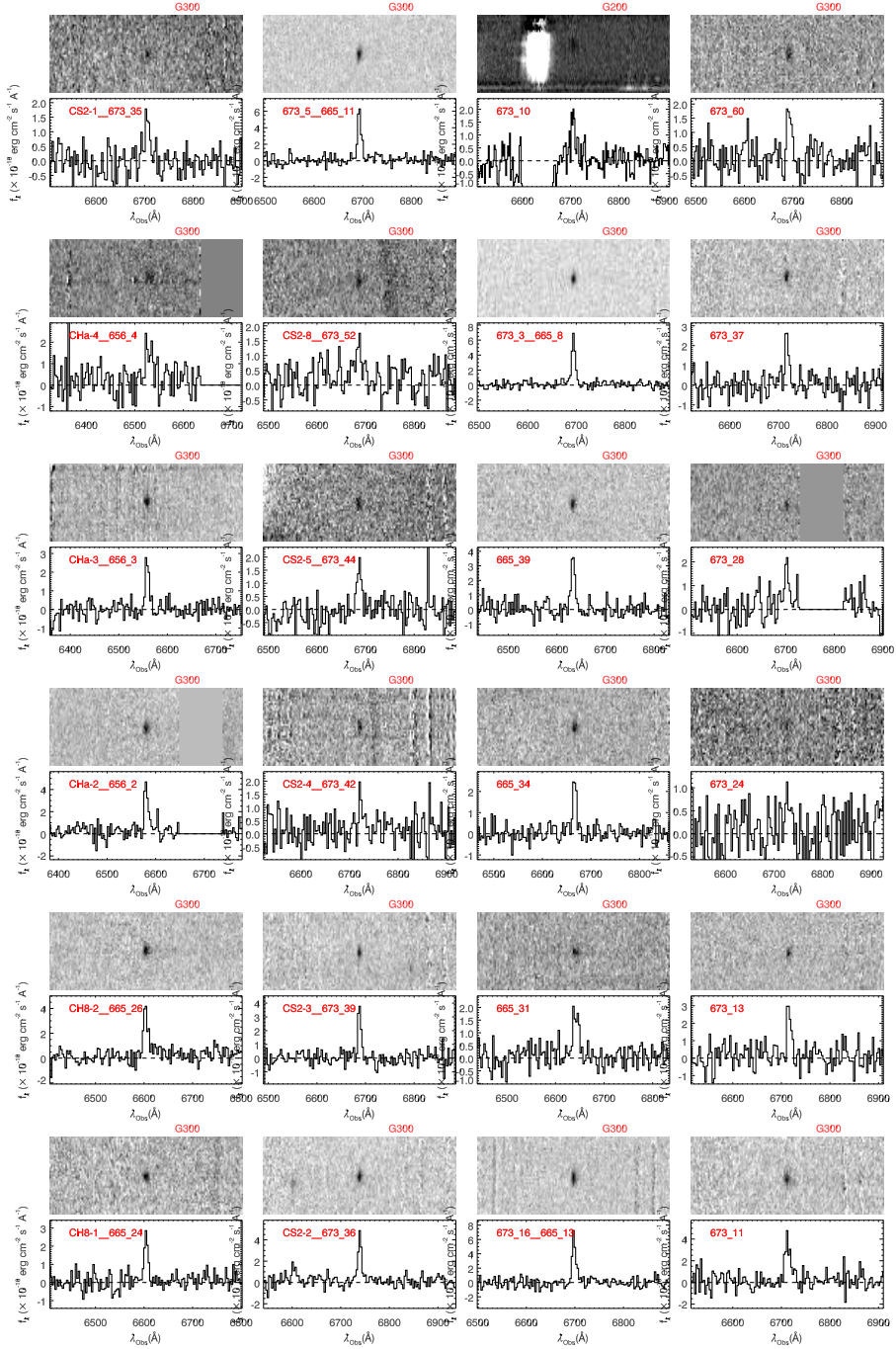} 
\caption{1D and 2D spectra of the spectroscopically confirmed LAEs (part 1). }
\label{conf1}
\end{center}
\end{figure*}

\begin{figure*}
\begin{center}
\includegraphics[totalheight=.9\textheight]{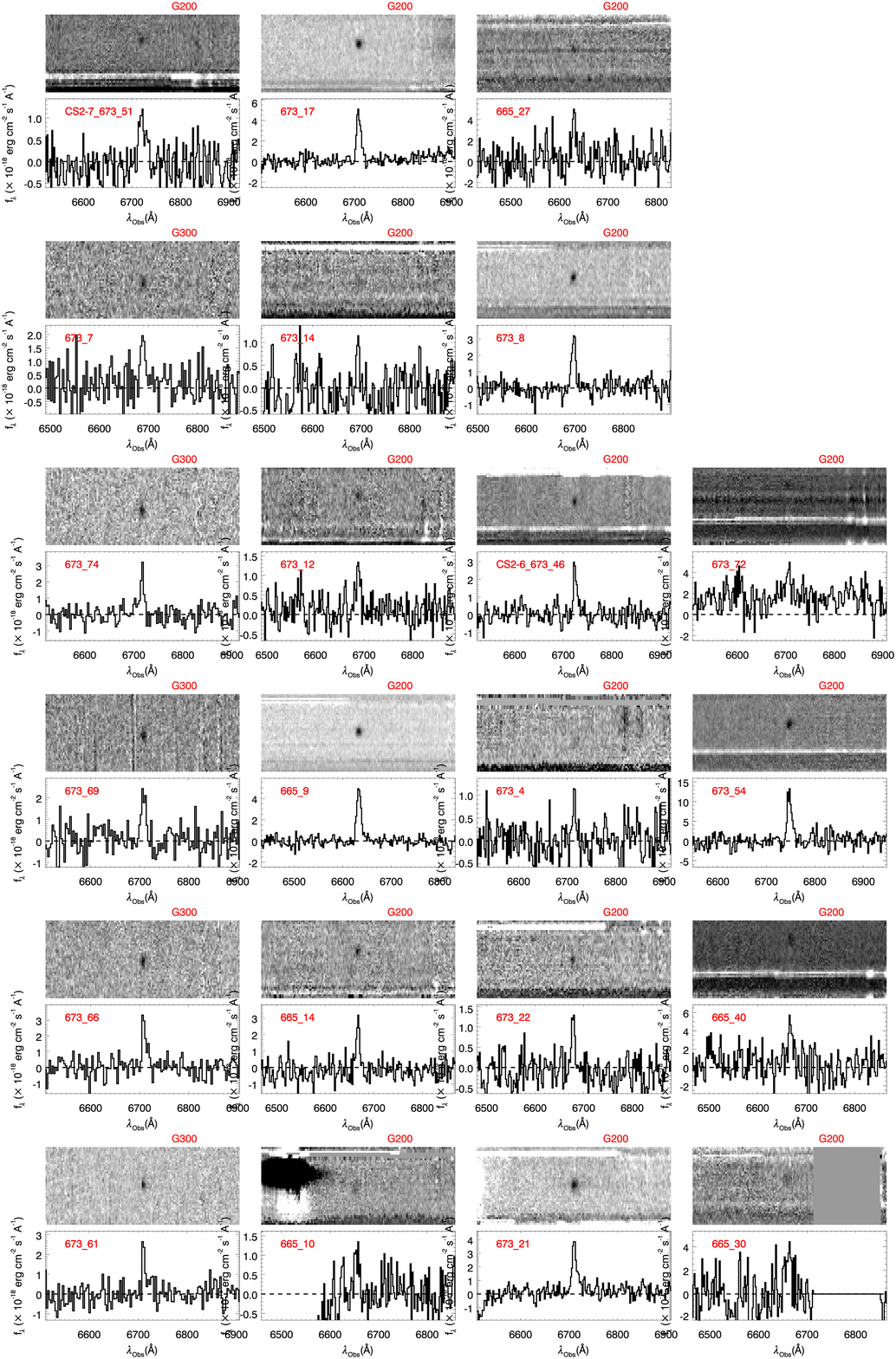} 
\caption{1D and 2D spectra of the spectroscopically confirmed LAEs (part 2). }
\label{conf2}
\end{center}
\end{figure*}

\begin{figure}
\begin{center}
\includegraphics[totalheight=0.3\textheight]{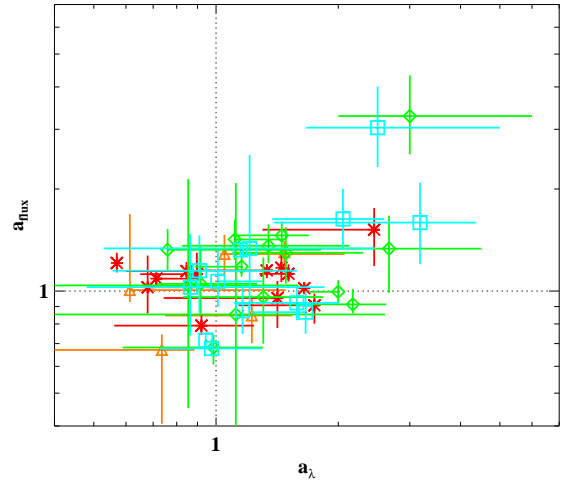} 
\caption{Two measurements of the line asymmetry for our 46 spectroscopically confirmed LAEs (see text for definitions of a$_\lambda$ and a$_f$.) The orange triangles and red stars are LAEs confirmed with the 10.2 ks and 12.2 ks 200 line/mm grism observations, and the green diamonds and cyan squares are LAEs confirmed with 16.2 ks and 14.4 ks 300 line/mm grism observations, respectively.}
\label{asym}
\end{center}
\end{figure}

\begin{figure}
\begin{center}
\includegraphics[totalheight=0.28\textheight]{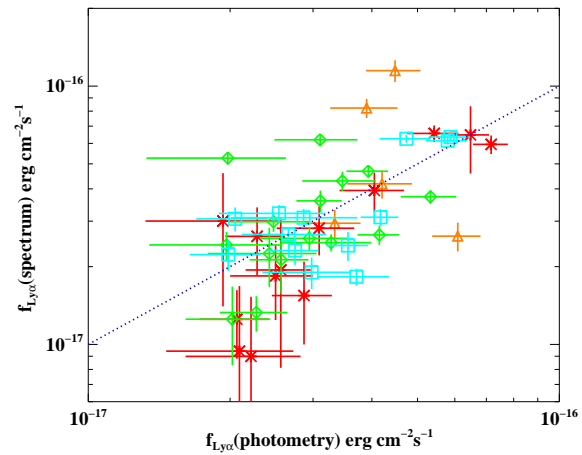} 
\caption{ The \lya\ photometric line flux (derived from the narrowband flux excess compared to the broadband) compared to the spectroscopic \lya\ line flux.  The orange triangles and red stars are LAEs confirmed with the 10.2 ks and 12.2 ks 200 line/mm grism observations, and the green diamonds and cyan squares are LAEs confirmed with 16.2 ks and 14.4 ks 300 line/mm grism observations, respectively.}
\label{fluxflux}
\end{center}
\end{figure}

\subsection{Spectroscopic Calibration}
\label{sec:spec:cali}

We observed one standard star in each mask with our LAE targets which we used to flux-calibrate the spectra.  The star is UID1147 (Pirzkal et al. 2005), and is well calibrated via HST grism spectroscopy (Pirzkal et al. 2005).   The comparison of \lya\ line fluxes integrated from spectroscopy and estimated from photometry is plotted in the figure \ref{fluxflux}. Here the \lya\ line flux from photometry is calculated as F$(Lya)_{Phot}$ = (f$_{NB,\lambda}$  - f$_{R,\lambda}$) / ($\frac{1}{FWHM_{NB}}$ - $\frac{1}{FWHM_R}$), and the redshift is estimated from the peak wavelength of the \lya\ line.
A large dispersion is seen in this plot, and the spectroscopic line flux from the 10.2 ks grism-200 mask are systematically  larger than that from the photometric data, possibly due to the slit-loss of the standard star observation. 
The slit width is 0.8\arcsec, and if the seeing is FWHM = 1\arcsec, then the average slit-loss is about 35.5\% (cf., 30.3\% with slit width of 1\arcsec\ and seeing 1\arcsec\ in  Kashikawa et al. 2011) if a LAE is unresolved and located in the center of a slit. In any case, the comparison between the line fluxes from the NB photometry and spectroscopy are less than a factor of two off in most cases. This is similar to the level of match from previous studies (Wang et al. 09, Kashikawa et al. 2011).  In the following analysis we use the photometric line flux.

\subsection{The stacked Spectrum}
\label{sec:spec:stack}

We follow Wang et al. (2009) to stack the spectra of the confirmed \lya\ emitters in order to better understand the nature of the LAEs. Excluding spectra with very large background noise at the estimated \lya\ line wavelengths, we stack 29 confirmed \lya\ emitters at z $\sim$ 4.5 with the grism 300 line/mm observation (see the blue dotted lines in Figure \ref{coadd}). For comparison, we also stack the 110 confirmed \lya\ emitters at z $\sim$ 4.5 in LALA Cetus field from Wang et al. 2009 (the magenta dotted lines in Figure \ref{coadd}), as well as a combined sample of all 139 confirmed \lya\ emitters at z $\sim$ 4.5 with the grism 300 line/mm observations (black solid lines in Figure \ref{coadd}). The redshifts are derived from their peak value of \lya\ line. We co-add the normalized spectra (normalized to f$_\lambda$(1216\AA) = 1) in a variance-weighted method, and using a 2$\sigma$ clipping algorithm (one iteration)  to remove artificial features(e.g., sky line residuals, CCD edges, etc.). The errors per wavelength unit are estimated from the standard deviation of the coadded spectra. 

\begin{figure}
\begin{center}
\includegraphics[totalheight=0.3\textheight]{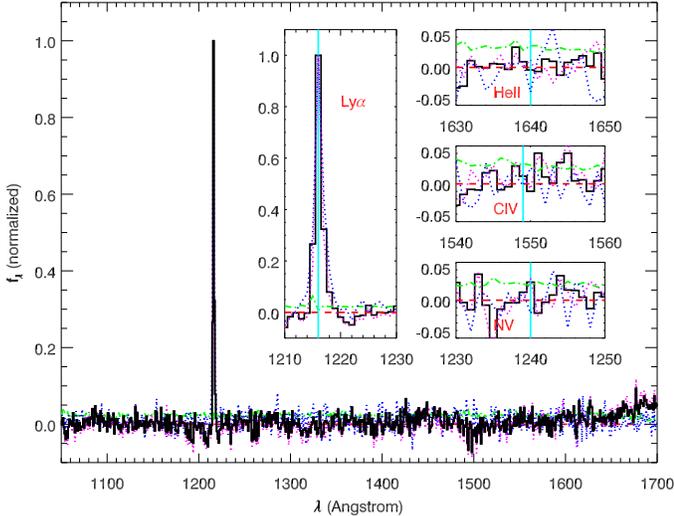} 
\caption{The stacked spectrum of 29 of our z $\sim$ 4.5 LAEs from the 300 line/mm grism observation (blue dotted lines).  We also show a stack of 110 z $\sim$ 4.5 LAEs from the LALA Cetus field (Wang et al. 2009; magenta dotted lines), as well as a combined stack of the two samples (139 LAEs, black solid lines). The redshifts are derived from their peak value of \lya\ line. The \lya\ line is highlighted in the cutout, and the co-added spectrum at the estimated position of C\,{\sc iv}, He\,{\sc ii} and N\,{\sc v} emission are also plotted. The 1-$\sigma$ errors of the co-added spectrum is shown as the green line.}
\label{coadd}
\end{center}
\end{figure}

The only visible line feature in the composite spectrum is the asymmetric \lya\ line (highlighted in the cutout of Figure \ref{coadd}). We measure a wavelength-based asymmetry of $a_\lambda$ = 1.7$^{+0.6}_{-0.7}$ and a flux-based asymmetry of $a_f$ = 1.2$^{+0.1}_{-0.1}$, consistent with those expected from the asymmetric profile of high-redshift \lya\ emission lines in previous works (e.g., Wang et al. 2009, Dawson et al. 2007).  The error bars on $a_\lambda$ and $a_f$ are estimated based on 1000 times Monte-Carlo simulations (similar to those we performed on the individual object spectra). 

There are no significant ($<$1$\sigma$) N\,{\sc v} $\lambda$ 1240 \AA, C\,{\sc iv} $\lambda$1549 \AA, or He\,{\sc ii} $\lambda$ 1640 \AA\ emission lines near the expected wavelengths (assuming that their line widths are equal to that of \lya) when coadding our 29 300 line/mm grism spectra. The resulting 2$\sigma$ upper limits on the line ratios are $f(\mbox{N\,{\sc v}}\lambda 1240)/f(\mbox{Ly-}\alpha ) <$ 4.3\%, $f(\mbox{C\,{\sc iv}}\lambda 1549)/f(\mbox{Ly-}\alpha ) <$ 6.0\% and $f(\mbox{He\,{\sc ii}}\lambda 1640)/f(\mbox{Ly-}\alpha ) <$ 4.4\%. Our constraint on C\,{\sc iv} is better than the previous study by Dawson et al. ($<$8\%; Dawson et al. 2004) based on Keck spectroscopy of 11 LAEs, but less than Wang et al. ($<$3.7\%; Wang et al. 2009) based on IMACS spectroscopy of 110 LAEs, though our constraint on $f(\mbox{He\,{\sc ii}}\lambda 1640)/f(\mbox{Ly-}\alpha )$ is stronger than that from Dawson et al. ($<$13\%) and Wang et al. ($<$7.4\%). Combining all the 139 LAE spectra which were taken with 300 line/mm grism with IMACS, we measure a 2-$\sigma$ upper limit of $f(\mbox{N\,{\sc v}}\lambda 1240)/f(\mbox{Ly-}\alpha ) <$ 6.7\%, $f(\mbox{C\,{\sc iv}}\lambda 1549)/f(\mbox{Ly-}\alpha ) <$ 7.0\% and $f(\mbox{He\,{\sc ii}}\lambda 1640)/f(\mbox{Ly-}\alpha ) <$ 5.9\%.

\section {DISCUSSION}
\label{sec:dis}

\subsection{AGN Fraction and Unobscured Star Formation Rate}
\label{sec:dis:agn}

In \citet{Zheng10}, we have previously performed an X-ray analysis on our LAE sample using the 2Ms CDF-S data \citep{Luo08} and the 250ks ECDFS data \citep{Lehmer05}. We found one detection in the ECDFS region, which has been spectroscopically confirmed by Treister et al. (2009) as a type 1 quasar (J033127.2-274247) at z = 4.48. X-ray stacking of the remaining sources shows a marginal detection (2.4$\sigma$). The current spectroscopic data show that about half of the signal was from one LBG and one possible LAE (NB673-27), both of which we reject from our spectroscopically confirmed LAE sample.  Excluding these two, we estimate a 3$\sigma$ constraint on the average X-ray luminosity of $L_{2-8 keV,rest}$ $<$ 2.4$\times$10$^{42}$ erg s$^{-1}$ \citep{Zheng10}.  In the new 4 Ms CDF-S data \citep{Xue11}, there are no individually detected LAEs.  A stacking analysis of the 4 Ms data shows that the average X-ray luminosity to $L_{2-8 keV, rest}$ $<$ 1.6$\times$10$^{42}$ erg s$^{-1}$ \citep{Zheng12}. Compared to the ratio of \lya\ to X-ray fluxes for typical AGNs\footnote{f$_{Ly\alpha}$/f$_{0.5-2keV}$ $\sim$ 1/8 for type 1 quasar template from \citet{Sazonov04}, and f$_{Ly\alpha}$/f$_{0.5-2keV}$ $\sim$ 1/4 for type 2 AGN like CXO 52 \citep[z = 3.288;][]{Stern02}, see \citet{Zheng10} and discussion therein.}, we can estimate that fewer than 2.1\% (4.2\%) of our LAEs could be high redshift type 1 (type 2) AGNs, and those hidden AGNs likely show low rest-frame \lya\ equivalent widths. 

We also search for other AGN indicators (e.g, the broad emission lines or high ionization emission line such as C\,{\sc iv}) in the spectra of our confirmed LAEs.  No such emission lines are visible in our spectra (note that among the 6 brightest LAE candidates, we only targeted and confirmed the 4th brightest object).  
We follow Dawson et al. (2007) and Wang et al. (2009) to use the  line ratio of C\,{\sc iv} and \lya\  from the composite optical spectra to constrain the upper limit of the AGN contamination fraction. 
Wang et al. (2009) showed that the C\,{\sc iv} line flux is invisible even after stacking 110 \lya\ emitters at z $\sim$ 4.5, for an upper limit of the C\,{\sc iv} to \lya\ line ratio of $<$ 3.7\% (2$\sigma$). In this work, we have presented the composite spectra of 29 \lya\ emitters at z$\sim$4.5 in Section 3.4. The  upper limit of our C\,{\sc iv} to \lya\ line ratio is  $<$ 6.0\% (2$\sigma$), more than  that of Wang et al. (2009). For comparison, a typical type II AGN has a C\,{\sc iv}/\lya\ ratio of 22\% (Ferland \& Osterbrock 1986). Similar AGN can therefore constitute no more than $\sim 6.0\%/22\%= 27\%$ of our \lya\ samples.  

Deep radio data is also an independent method to search for the presence of AGNs in our LAEs. The entire ECDFS has been imaged with the VLA to a typical sensitivity  of 7.2 $\mu$Jy per 2.8\arcsec $\times$ 1.6\arcsec\ beam (1-$\sigma$; Miller et al. 2008, Miller et al. in Prep.). The radio catalogue from Miller et al. was searched with a match radius of 3\arcsec\ with our LAE candidates and two objects were found. One object (J033127.2-274247, with a separation of $<$1\arcsec) is the X-ray detected and spectroscopically confirmed z=4.48 quasar. The other one (NB673\_21) with a separation of 2.4\arcsec\ to the radio detection, is contaminated by other nearby sources within 3\arcsec\ in the optical B-band image, thus the radio signal is likely not from our LAE. 
A clipped stack of all 112 candidates (excluding the three highest and three lowest measurements at each pixel, and so excluding the two individually detected sources) reveals no composite detection,  down to 0.67 uJy rms ($\sim$ 7.2/sqrt(112-6)).   
The radio-loud fraction of quasars are thought to be a strong function of redshift and optical luminosity, and at z $\sim$ 4 the fraction is $\sim$ 3.5$\pm$2.5\% with -27.4 $<$ M$_{2500}$ $<$ -25.0 (Jiang et al. 2007), implying that there might 4$^{+3}_{-3}$ radio-quite quasars in our LAE sample based on the UV luminosities of our objects. However,  except for the one radio-loud quasar confirmed through other means (X-ray, radio and deep spectroscopy), we did not find any other AGNs (e.g., the existence of C\,{\sc iv}, He\,{\sc ii} emission lines, or broad Lya line) in the 46 confirmed LAEs' spectra. 

The star-formation activity in star-forming galaxies also contributes X-ray and radio emission, which are both useful tracers as they are unobscured by dust (Zheng et al. 2012). Although the X-ray and radio radiation from normal galaxies are much fainter than that from AGNs, the number density of galaxies is significantly larger than that of AGNs. Xue et al. (2011)  pointed out that in the 4Ms CDF-S region, the X-ray radiation from galaxies begins to dominate that of AGNs at the faintest detectable flux levels. If the X-ray radiation is all due  to star formation, our X-ray average flux would correspond to a 2$\sigma$ upper limit of unobscured star-formation rate (SFR) $<$214 M$_\odot$   yr$^{-1}$ (Ranalli et al. 2003). If the radio fluxes are converted into star formation rate using the conversion rate of Yun et al. (2001), the rms sensitivity of the radio map corresponds to upper limit of SFR $\lesssim$1700 M$_\odot$/yr at z$\sim$4.5.  The radio stacking of our LAE candidates can be translated to a 2-$\sigma$ upper limit of SFR $\lesssim$100M$_\odot$/yr at z$\sim$4.5. Since the sensitivity of the X-ray image is non-uniform, the ratio stacking gives better constraint than X-ray stacking.  The average SFR from the \lya\ emission line (with the relation from Kennicutt 1998, under Case B recombination from Brocklehurst 1971) is about 5 M$_{\odot}$ yr$^{-1}$, though the resonantly scattering nature of \lya\ photons renders this measurement very uncertain. If we assume that the SFR from X-ray or radio is consistent with the intrinsic SFR, the ratio of SFRs from observed \lya\ and from upper limits of X-ray or radio can be used to constrain the lower limit of \lya\ escape fraction, which is $f_{esc}(Ly\alpha)$ $>$2.4-5.0\% for z $\sim$ 4.5 LAEs at the 95.4\% confidence level. This limit is consistent with the average escape fraction from various methods at lower redshift, e.g., an average fraction of 29\% at 1.9 $\leq$ z $\leq$ 3.5 from \citep{Blanc11} with the comparison between \lya\ flux and dust-corrected UV continuum, $>$ 7\% (2 $\sigma$ upper limit) at z = 2.1 and 3.1 from Zheng et al. (2012) with the comparison between \lya\ flux and X-ray flux, and $\sim$12\%--30\% at z= 2.2 from Nakajima et al. (2012) with the comparison between \lya\ and H$\alpha$ luminosity.  Larger samples at z = 4.5 or deeper X-ray or radio data are needed to give better constraints on the \lya\ escape fraction.
 
\subsection{Population III stars}
\label{sec:dis:pop3}

Population III (Pop III) stars are thought to have very strong He\,{\sc ii} emission (He\,{\sc ii} $\lambda$ 1640 line, the Balmer $\alpha$ transition of singly ionized helium). In this work, we find an upper limit of the He\,{\sc ii}-to-\lya\ ratio of $<$ 4.4\% (2-$\sigma$ upper limit), lower than the $<$7.4\% of Wang et al. (2009). Schaerer (2002) derived the relation between He\,{\sc ii} recombination line luminosity and the SFR$_{PopIII}$ under a constant star formation,
\begin{eqnarray}
L_{He\,{\sc ii}} & = & c_{1640} \times (1-f_{esc})  \times Q(He^+)\times\big(\frac{\hbox{SFR}_{PopIII}}{M_\odot yr^{-1}}\big) \\ \nonumber
 & = & L_{1640,norm} \times \big(\frac{\hbox{SFR}_{PopIII}}{M_\odot yr^{-1}}\big),
\end{eqnarray}
where c$_{1640}$ is the He\,{\sc ii} $\lambda$1640 emission coefficient given in Table 1 of Schaerer (2003): c$_{1640}$ = 5.67$\times$10$^{-12}$ erg for $T_e$ = 30$k$ K,  $Q(He^+)$ is the number of He$^+$ ionizing photons per second, f$_{esc}$ is the fraction of total ionizing radiation released into the IGM without being coupled to the ISM in the galaxy, and $L_{1640,norm}$ is the theoretical He\,{\sc ii} $\lambda$1640 line luminosity normalized to SFR = 1 M$_{\odot} yr^{-1}$. Here we assume $f_{esc}$ = 0, a Salpeter IMF of 50--500 M$_\odot$, and no mass loss, and we derive a 2-$\sigma$ upper limit on SFR$_{PopIII}$ $<$ 0.30 M$_\odot$ yr$^{-1}$. If we take average SFR$_{Lya}$ $\sim$ 5 M$_\odot$ yr$^{-1}$ and \lya\ escape fraction of 29\% (Blanc et al. 2011), the ratio of SFR$_{PopIII}$ to SFR$_{total}$ (SFR$_{obs}$) at z$\sim$ 4.5 is $<$ 1.74\% (6\%) at 95.4\% confidence (2-$\sigma$) level. Since \lya\ photons are resonantly scattered and easily affected by velocity and geometry of ISM, we should use  the UV continuum and dust properties to get more accurate estimates on the SFR$_{obs}$ and SFR$_{total}$. Converted from GOODS F775W $i'$ band flux and dust extinction from SED fittings of our 14 LAEs (Finkelstein et al. 2009a), we got average SFR$_{UV}$ $\sim$ 24 M$_\odot$ yr$^{-1}$ and dust-corrected SFR$_{UV,dust-corr}$ $\sim$ 100 M$_\odot$ yr$^{-1}$ (see table 3 of Zheng et al. 2012). So the ratio of SFR$_{PopIII}$ to SFR$_{total}$ (SFR$_{obs}$) at z$\sim$ 4.5 is $<$ 0.3\% (1.25\%) at 95.4\% confidence (2-$\sigma$) level. This implies that the Pop III stars at z $\sim$ 4.5 are very rare, consistent with results from simulations (e.g., Scannapieco et al. 2003).

Recently, McLinden et al. (2011), Finkelstein et al. (2011) and Hashimoto et al. (2012) reported the discovery of velocity offsets in LAEs at z$\sim$2--3.1, in which the \lya\ emission had a slightly higher redshift (likely due to outflows with velocity $\sim$ 150--300 km/s) than the rest-frame optical [O\,{\sc iii}] or H$\alpha$ emission. In the co-adding process above, we fix the \lya\ peak as the systemic redshift for each LAEs. If outflows (or some other kinematic process) is shifting the redshift of the \lya\ line, the above stacking analysis may not be applicable.  Thus one must use caution when interpreting results from emission lines other than \lya\ based on our stacked LAE spectra.

\subsection{Ly$\alpha$ Luminosity Function}
\label{sec:lyalf}
The \lya\ luminosity functions are fundamental observational quantities of LAEs. In this section, we introduce the method used to measure the \lya\ luminosity function in our sample of z $\sim$ 4.5 LAEs in the ECDFS (Section \ref{sec:lyalf:z45lyalf}).  We then compare our \lya\ luminosity functions with two others from LAE surveys in the LALA-Bootes (Dawson et al. 2007) and LALA-Cetus (Wang et al. 2009) fields at the same redshift (Section \ref{sec:lyalf:lyalfz45}).  Finally, we explore the evolution of the \lya\ luminosity function by comparing \lya\ luminosity functions from different LAE surveys at different redshifts (section \ref{sec:lyalf:evolution}). In Section \ref{sec:lyalf:evolution}, we also study the evolution of the global \lya\ escape fraction, which is defined as the ratio of the star-formation rate density derived from LAEs' \lya\ luminosity functions to the star-formation rate density from dust-corrected rest-frame UV luminosity functions.

\subsubsection{Ly$\alpha$ Luminosity Function at z$\sim$ 4.5 in CDF-S}
\label{sec:lyalf:z45lyalf}

\begin{figure}
\begin{center}
\includegraphics[totalheight=0.3\textheight]{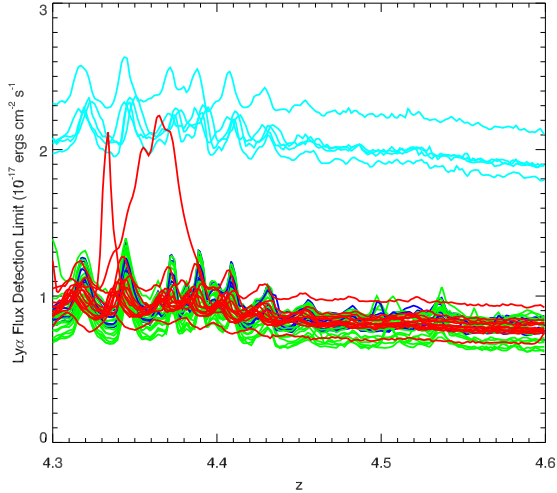} 
\caption{The spectral flux limit as a function of redshift estimated by assuming the \lya\ 1216\AA\ emission line is shifted to the observed wavelength. Here cyan and red lines are the 3-$\sigma$ upper limits from the 10.2 ks and 12.2 ks exposure with 200 line/mm grism observation, and the blue and green lines are for 16.2 ks and 14.4 ks exposure with 300 line/mm grism, respectively.  We assume the \lya\ line has a width of FWHM = 5\AA.} 
\label{speclimit}
\end{center}
\end{figure}

\begin{figure}
\begin{center}
\includegraphics[totalheight=0.3\textheight]{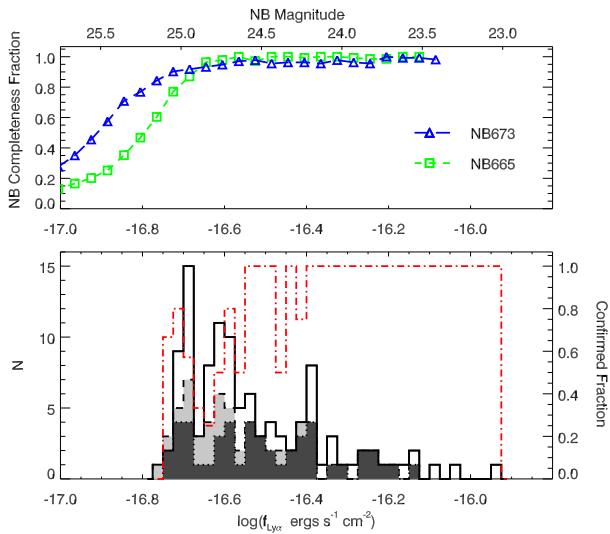} 
\caption{Top: Narrowband completeness fraction vs. narrowband magnitude (top x-axis), and the fluxes on the middle x-axis are converted from narrowband magnitudes only; Bottom: Distribution of the \lya\ fluxes ( continuum-corrected) for all candidate LAEs (empty histogram), and those targeted (light grey histogram) and spectroscopically confirmed (dark grey histogram) LAEs. The dot-dashed line shows the spectroscopic successful fraction (i.e., the ratio of the confirmed number to the targeted number per flux bin). } 
\label{frac}
\end{center}
\end{figure}

\begin{figure}
\begin{center}
\includegraphics[totalheight=0.3\textheight]{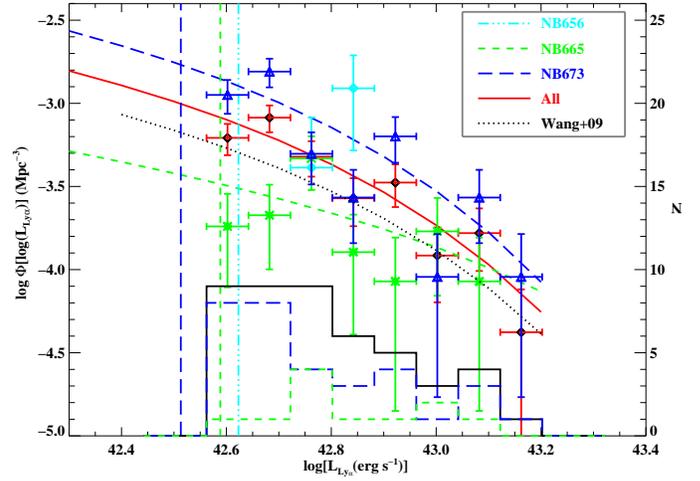} 
\caption{Differential \lya\ luminosity function for our z = 4.5 sample, computed using the $V/V_{max}$ method. The error bars are the 1 $\sigma$ statistical uncertainties given by the root variance of the quantity $1/V_{max}$. The black solid histogram gives the number of individual sources contributing to each luminosity bin. The dotted curve shows the best-fit model from Wang et al. 2009 with a fixed faint-end slope $\alpha$ = -1.5. The cyan, green and blue points are presented for our narrowband NB656, NB665 and NB673 data, respectively. The vertical color lines show the detection limits in the corresponding narrowband, and the colored curves show our fitted Schechter functions with a fixed faint-end slope $\alpha$ = -1.5 for the corresponding narrowband data.} 
\label{lyalf}
\end{center}
\end{figure}

\begin{figure}
\begin{center}
\includegraphics[totalheight=0.3\textheight]{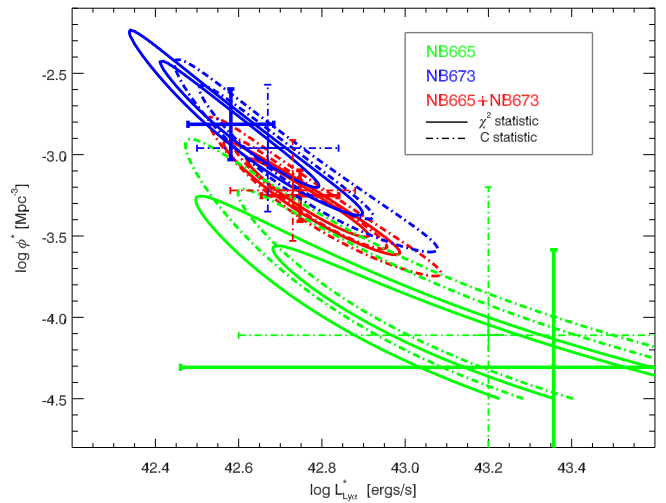} 
\caption{Contour plot of the parameters L$^*$ and $\Phi^*$ for our z = 4.5 sample. Here green, blue and red contours are the results for the NB665, NB673, and NB665+NB673 samples, respectively, with the two contours denoting the confidence level of 68\% and 90\% ($\Delta \chi^2$ = 2.3 and 4.6, or $\Delta C$  = 2.3 and 4.6). The solid and dash-dotted lines are fitting results with $\chi^2$ statistic and Cash $C$-statistic, respectively. Note that all best-fit parameters are obtained by fixing $\alpha$ = -1.5. } 
\label{lyalffitz45each}
\end{center}
\end{figure}

\begin{figure}
\begin{center}
\includegraphics[totalheight=0.3\textheight]{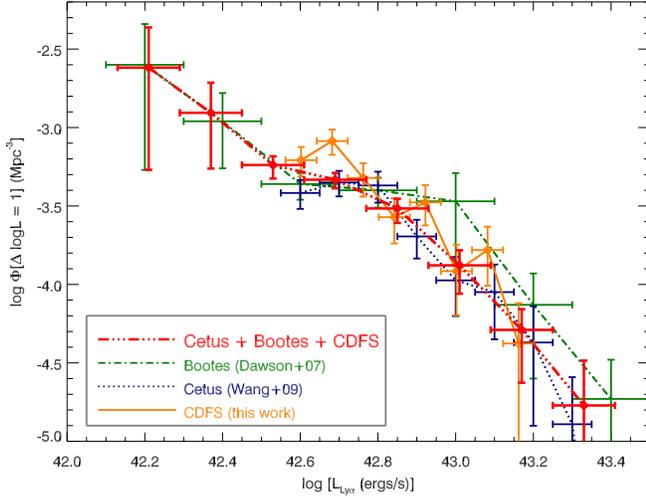} 
\caption{Differential \lya\ luminosity function for all z = 4.5 samples, including the data from CDF-S (orange symbols, this work), LALA-Bootes \citep[dark-green symbols][]{Dawson07}, LALA-Cetus \citep[navy symbols,][]{Wang09}, and the three combined fields (red symbols, this work).} 
\label{lyalfz45}
\end{center}
\end{figure}

\subsubsection*{ \textbf{-- V$_{max}$ method}}

We now present the \lya\ luminosity function based on our spectroscopically confirmed LAE sample at z $\sim$ 4.5 in the ECDFS. We choose a modified version of the $V/V_{max}$ method (Dawson et al. 2007). The comoving volume $V_{max}$ for each confirmed LAE was calculated, where $V_{max}$ is the volume where the source could be selected by our survey. The survey area is 34\arcmin $\times$ 33\arcmin\ from z =  4.44 to z = 4.56 (NB665 and NB673 filters) and 160 square arcmins from z = 4.36 to z = 4.43 (NB656 filter; the NB656 image was shallower than the others, and we therefore selected objects that were detected in the GOODS ACS images and that had a narrowband excess; see Finkelstein et al 2008 for more detail). Due to the limited redshift range of our survey, objects with the same luminosities only show a decrease of 0.015 dex on their line fluxes in the NB665 and NB673 images, thus $V_{max}$ estimated here is nearly equal to our survey volume. Note that we assume that the shape of our narrowband filters are top-hat, so we do not account for the filter transmission effects in the luminosity function calculation. We use the formulae below to measure our \lya\ luminosity function:
\begin{equation}
\Phi(L)d L = \sum_{L-d L/2 \le Li \leq L+d L/2} \frac{1}{V_{max}(L_i)\times f_{comp}(L_i)},
\end{equation}
here V$_{max}(L_i)$ is the maximum volume for LAEs with \lya\ luminosity L$_i$ which can be found in our narrowband surveys, and $f_{comp}(L_i)$ is the completeness fraction for LAEs with \lya\ luminosity L$_i$.
Dawson et al. (2007) had considered two types of incompleteness in their \lya\ luminosity function measurement: the target incompleteness ($f_{target-comp}(L_i)$, i.e., not all candidates were targeted) and the spectroscopic sensitivity depth $f_{spec-comp}(L_i)$. The target incompleteness is the ratio of the number of all candidate LAEs to the number of those targeted for spectroscopy in relative \lya\ flux bins (see figure \ref{frac}). We ignore the spectroscopic sensitivity incompleteness ($f_{spec-comp}(L_i)$ = 1),  as our spectra are generally more sensitive than our narrowband photometry (see figure \ref{speclimit} and \ref{fluxflux}). 
We take into account the incompleteness of our narrowband images ($f_{nb-comp}(L_i)$) in our \lya\ luminosity function calculation.  The narrowband incompleteness was estimated following Hibon et al. (2010): we added 200 artificial star-like objects (LAEs are point sources in these NB images) per bin of 0.1 mag to the NB673 image, then ran SExtractor on this image for object detection (similar to that done during the LAE selection process), and obtained a direct measure of the narrowband completeness fraction by counting the number of artificial stars detected in each magnitude bin over the total input artificial objects. The 90\% (50\%) completeness level of our narrowband NB665 and NB673 are 24.9 and 25.1 (25.2 and 25.5), respectively. When connecting narrowband completeness fraction and corresponding luminosity, we assume zero continuum here. The target incompleteness and the narrowband incompleteness are combined together in calculating the luminosity function (see figure \ref{frac}). To be conservative, we assume that targets not detected in our spectroscopic data are not \lya\ galaxies. Thus, we obtain a final completeness fraction of $f_{comp}(L_i)$ =$f_{target-comp}(L_i)$ $\times$ $f_{nb-comp}(L_i)$.

\subsubsection*{ \textbf{-- Schechter function fitting, $\chi^2$ statistic, and \hbox{Cash} C-statistic}}

Our derived \lya\ luminosity function of LAEs at z$\sim$ 4.5 in the ECDFS is shown in Figure \ref{lyalf}. Following Malhotra \& Rhoads (2004), we fit the \lya\ luminosity function with a Schechter function
\begin{equation}
\Phi(L)d L = \frac{\Phi^*}{L^*}\left (\frac{L}{L^*}\right )^\alpha \exp\left (-\frac{L}{L^*}\right ) d L .
\end{equation}
We use the IDL program \textit{mpfit} to fit our \lya\ luminosity function data with a Schechter function using $\chi^2$ statistics ($\chi^2$ = $\sum_{i=1}^{N}(\Phi_i-\Phi_{mod})^2/Err_\Phi^2$). We did not consider the photometric errors of luminosities in the fitting, as we divided our sample into bin size of 0.08 dex, which corresponds to a $\sim$5-$\sigma$ detection in the faintest L bin. Thus photometric error will primarily affect our faintest bin, which directly affects the constraint on the faint-end slope.  However, given the results from previous studies, are data are likely not deep enough to robustly constrain $\alpha$.  Thus, following previous studies, we choose to fix the faint-end slope $\alpha$ = -1.5. 
We find best-fit parameters of log$_{10}$($L^*$) = 42.75 $\pm$0.09 and log$_{10}$($\Phi^*$) = -3.25$\pm$0.16 ($\chi^2/dof$ = 6.54/6).  When performing the same \lya\ luminosity function measurements on the samples from the two narrowband images seperately, the best-fit parameters are log$_{10}$($L^*$) = 43.36 $\pm$0.90 and  log$_{10}$($\Phi^*$) = -4.31 $\pm$0.72 ($\chi^2/dof$  = 4.15/5) for the NB665 sample, and log$_{10}$($L^*$) = 42.58 $\pm$0.10 and  log$_{10}$($\Phi^*$) = -2.81 $\pm$0.22 ($\chi^2/dof$ = 13.88/6) for NB673 sample. The number of confirmed LAEs in the two narrowband images are too few (especially in NB665) to constrain well both L$^*$ and $\Phi^*$ (see figure \ref{lyalffitz45each}).  We thus fix L$^*$ to the best-fit value from the fit for all confirmed LAEs in our sample (log$_{10}$($L^*$) = 42.75), and we find that $\Phi^*$ = -3.68$\pm$0.11 ($\chi^2/dof$ = 5.7/6) in NB665 and  $\Phi^*$ = -3.13$\pm$0.06 ($\chi^2/dof$ = 15.5/7) in NB673.  The difference in the characteristic number density $\Phi^*$ is different in the two redshift bins at greater than 3$\sigma$ significance.    

However, the above fitting method assumed that uncertainties in the number counts are Gaussian, while they in reality are likely Poissonian.  At large numbers the difference is negligible; however, many of our luminosity bins have relatively few objects, thus we re-measure our luminosity functions using the Cash $C$-statistic (Cash 1979): C = -2 ln$L$ = -2$\sum_{i=1}^N (n_a \hbox{ln}e_a-e_a-\hbox{ln}n_a!)$ = 2$\sum_{i=1}^N e_a - n_a + n_a \times (\hbox{ln}n_a - \hbox{ln}e_a)$, which is appropriate when uncertainties are Poissonian. Here n$_a$ and e$_a$ are, respectively, are the observed and expected number in sample, and N is the total number of samples. In this section, N is the number of luminosity bins from the $V_{max}$ method, and n$_a$ and e$_a$ are, respectively, the observed and expected number of LAEs in each luminosity bin. From the observed $\Phi(L)\hbox{d}L$ and n$_a(L)$, we can estimate the average luminosity density per galaxy, as $\Phi(L)\hbox{d}L$ / n$_a(L)$. So using a simple grid of parameters $L^*$ and $\Phi^*$, the expected number is e$_a$$(L)$ = $\Phi(L^*,\Phi^*)\hbox{d}L$ / ($\Phi(L)\hbox{d}L$ / n$_a(L)$). The resulting luminosity functions are plotted in Figure \ref{lyalffitz45each} and Table \ref{statisfit}. 

\begin{table*}
\begin{minipage}{100mm}
\caption{The Schechter function fitting results of the \lya\ luminosity functions at z$\sim$ 4.5 in our two narrowband fields. Note that we fixed the faint-end slope to $\alpha$ = -1.5 in the fitting.}
\begin{tabular}{@{}lcccccc@{}}
\hline \hline
Field 		& dof & Statistic 		&    L$_*$ & $\Phi^*$ & Covar & L$_*$$\Phi^*$  \\
&    &  &   log$_{10}$(erg/s) & log$_{10}$(Mpc$^{-3}$) &  & log$_{10}$ (erg/s Mpc$^{-3}$) \\
\hline
NB665 		&	5 &	$\chi^2$ = 4.1 & 	 43.36$\pm$0.90 	& -4.31$\pm$0.72 	& $ \Bigl(\begin{array}{cc} 0.804 & -0.641 \\ -0.641 & 0.522 \end{array} \Bigr)$ 	 &  39.05$\pm$0.21\\
			&	   &   $C$-stat =  2.6  & 	  43.2$\pm$0.60	&	-4.11$\pm$0.91	&	$ \Bigl(\begin{array}{cc} 0.360 & -0.581 \\ -0.581 & 0.828 \end{array} \Bigr)$
							&  39.09$\pm$0.16	\\ \hline     
NB673		&	6 &	$\chi^2$ = 13.9 &  42.58$\pm$0.10	 & -2.81$\pm$0.21	 &$ \Bigl(\begin{array}{cc} 0.0106 & -0.0217 \\ -0.0217 & 0.0472 \end{array} \Bigr)$ &  39.77$\pm$0.12 	\\
			&	   &   $C$-stat =  8.0  & 	  42.67$\pm$0.17		&	-2.96$\pm$0.39		&	$ \Bigl(\begin{array}{cc} 0.029 & -0.087 \\ -0.087 & 0.152 \end{array} \Bigr)$					&		39.71$\pm$0.08		\\ \hline    
NB665+NB673	 &	6 &	$\chi^2$ = 6.5 &  42.75$\pm$0.09 	& -3.25$\pm$0.16	 &  $ \Bigl(\begin{array}{cc} 0.0086 & -0.014 \\ -0.014 & 0.025\end{array} \Bigr)$ & 39.50$\pm$0.07	\\
			&	   &   $C$-stat =  3.3  &   42.73$\pm$0.15	&	-3.22$\pm$0.31	&	$ \Bigl(\begin{array}{cc} 0.023 & -0.057 \\ -0.057 & 0.096 \end{array} \Bigr)$							&	39.51$\pm$0.07			\\ \hline    
 \hline
\label{statisfit}
\end{tabular}
\footnotetext[1]{The error of $L^*\Phi^*$ is calculated from the covariance matrix, as var( lgL$_*$+log$_{10}\Phi^*$ ) = var(lgL$_*$) + var(log$_{10}\Phi^*$) + 2cov(lgL$_*$,log$_{10}\Phi^*$).  }
\end{minipage}
\end{table*}

The best-fit parameters of $L^*$ and $\Phi^*$ are similar from both the $\chi^2$ statistic and the $C$-statistic in the combined sample from both narrowband images, while the fitting results in individual narrowband samples show better behavior with the $C$-statistic (smaller errors and less separation) than that with $\chi^2$ statistic, seemingly confirming that luminosity function fitting with the $C$-statistic is more appropriate when the source numbers in each luminosity bin are small (n$_a$ $\lesssim$ 2), and that the fitting method is irrelevant when the source numbers are large enough (n$_a$ $\gtrsim$ 4) in each luminosity bin. In the following sections, we only use the $\chi^2$ statistic when comparing luminosity functions from the literature, as in a number of cases on the number densities were published (and the $C$-statistic requires the actual numbers). 


The contours in Figure \ref{lyalffitz45each}  also highlight that the parameters $L^*$ and $\Phi^*$ are correlated.  It thus becomes interesting to examine the quantity $L^*\Phi^*$, which is proportional to the integrated \lya\ luminosity density.  One direction of the covariance matrix corresponds to the uncertainty in this quantity (or lg($L^*$)+lg($\Phi^*$) in the log-normal form), thus we require the covariance matrix of the luminosity function to calculate the errors of $L^*\Phi^*$, as var( lgL$_*$+log$_{10}\Phi^*$ ) = var(lgL$_*$) + var(log$_{10}\Phi^*$) + 2cov(lgL$_*$,log$_{10}\Phi^*$). The errors of  $L^*$, $\Phi^*$, and $L^*\Phi^*$, as well as the covariance matrix are presented in Table \ref{statisfit}. Interestingly, even though $L^*$ and $\Phi^*$ vary with different statistics in the fitting in the same field, the value of $L^*\Phi^*$ is not changed.

\subsubsection*{ \textbf{-- Cosmic variance}}

The resulting luminosity functions with both the $C$-statistic and $\chi^2$ statistic in each narrowband filter show a difference of $\gtrsim$90\%. This can be due to systematic errors such as selection effects, observational depth, or field to field variation. From Figure 1 we see that most bright LAEs are confirmed, and via \lya\ equivalent width simulations (Zheng et al. in prep.), we find that the EW lower limit imposed by our selection does not significantly affect the properties of the sample. Thus any selection effects likely only affect the faint end of our luminosity function, which has little effect when we fix the faint-end slope of $\alpha$ = -1.5. The observational depth also primarily affects the faint end of our luminosity function. Thus, the difference of the resulting luminosity functions in our two narrowband fields is primarily the field-to-field variation, with a significance level of $\gtrsim$90\%.

This difference can be due to cosmic variance, though some amount is also due to the Poisson uncertainty.  We examine the impace of the former via the Cosmic variance calculator by Trenti \& Stiavelli (2008), finding the total fractional error on number counts for our survey of 37.2\% (1-$\sigma$), in which Cosmic variance composes the majority of the uncertainty of 33.7\%, and the Poisson uncertainty accounts for 15.8\% (survey parameters: area = 30 x 34, mean redshift = 4.5, and redshift interval = 0.0658; catalog parameters: intrinsic number of objects = 80, Halo filling factor = 1.0, and completeness = 0.5), and the simulated observed number counts is 56$\pm$20 in one field. Our photometric data find 33 and 75 LAE candidates in NB665 and NB673, respectively, and spectroscopic observations confirmed 11 of 17 in NB665  and 32 of 44 in NB673. Correcting for the spectroscopic successful fraction (see figure \ref{frac}), we estimate that there are 24 and 55 LAEs in our NB665 and NB673 fields, respectively, corresponding to a fractional error of 39.2\%. This  is slightly larger than the estimated total fractional error (37.2\%), implying the existence of cosmic variance in our data at $\geq$ 1-$\sigma$ level.

  \subsubsection{Unified Ly$\alpha$ Luminosity Function at z $\sim$ 4.5}
  \label{sec:lyalf:lyalfz45}

We combine all the spectroscopically confirmed LAEs at z $\sim$ 4.5, from the LALA Bootes field (Dawson et al. 2004, 2007), the LALA Cetus field (Wang et al. 2009), and this work on the ECDFS, to derive a unified Ly$\alpha$ luminosity function at z $\sim$ 4.5 (see Figure \ref{lyalfz45}). When fitting the Schechter function, we fix the faint-end slope of $\alpha$ = -1.5 and use the $\chi^2$ statistic (for ease of comparison with the literature). The results are shown in Table \ref{comblyalf} and Figure \ref{lyalffitz45}.  The best-fit parameters are log$_{10}$($L^*$) = 42.83 $\pm$0.06 and  log$_{10}$($\Phi^*$) = -3.48 $\pm$0.09 ($\chi^2/dof$ = 2.9/6) for the unified z $\sim$ 4.5 LAE sample.  The best-fit parameters of L$^*$ and $\Phi^*$ of the combined z$\sim$ 4.5 LAEs are located within the 90\% confidence level of the L$_*$ and $\Phi^*$ contours of the three separate fields. Considering the nearly same selection method and observational depths of the three fields, any differences in  L$_*$ and $\Phi^*$ of the three separate fields is likely at least partially due to cosmic variance, though as we showed above, the Poissonian uncertainties play a strong role.
  

\begin{figure}
\begin{center}
\includegraphics[totalheight=0.3\textheight]{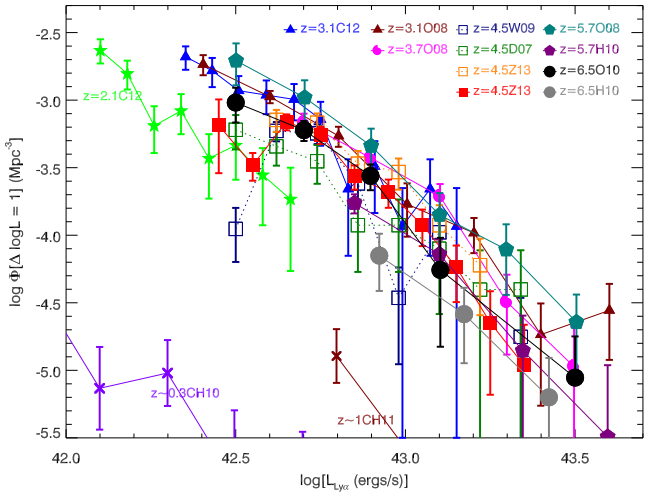} 
\caption{We compare differential \lya\ luminosity functions from various surveys at different redshifts, including 2 GALEX selected LAE samples at  z $\sim$ 0.3 (Cowie et al. 2010, purple line) and $\sim$ 1.0 (Cowie et al. 2011, dark-red line), and all z $>$ 2 narrowband selected LAEs at redshifts z $\approx$ 2.1, 3.1, 3.7, 4.5, 5.7, and 6.5.  Differential \lya\ luminosity functions for the combined sample of all z $\approx$ 4.5 surveys, including LALA Bootes (dark green squares, Dawson et al. 2007), LALA Cetus (navy squares, Wang et al. 2009) and CDF-S (orange squares, this work), where our combined sample is denoted in red filled squares.  Other symbols are results from the literature at z $\approx$ 2.1 (Ciardullo et al. 2012, dot-dashed green stars), z $\approx$ 3.1 (blue triangles for Ciardullo et al. 2012,  and brown triangles for Ouchi et al. 2008), z $\approx$ 3.7 (Ouchi et al. 2008, magenta spirals),  z $\approx$ 5.7 (dark-cyan pentagons for Ouchi et al. 2010, and dark-magenta pentagons for Hu et al. 2010), and z $\approx$ 6.5 (black circles for Ouchi et al. 2010, and grey circles for Hu et al. 2010).}
\label{difflyalf}
\end{center}
\end{figure}

\begin{figure}
\begin{center}
\includegraphics[totalheight=0.3\textheight]{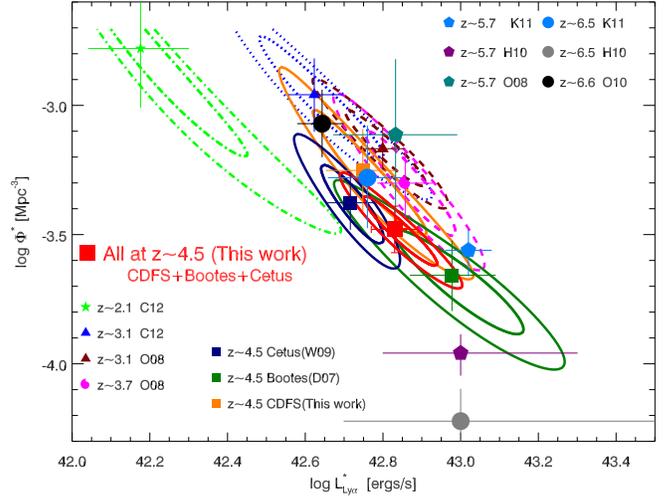} 
\caption{Confidence intervals of the best fit parameters of L$^*$ and $\Phi^*$ on the differential \lya\ luminosity functions  from various surveys at different redshifts. The samples include all z $\sim$ 4.5 surveys: LALA Bootes (dark green lines, Dawson et al. 2007), LALA Cetus (navy lines, Wang et al. 2009), and CDF-S (orange lines, this work), and all z $\sim$ 4.5 combined sample denoted in red.  Other contours are results from the literature at z $\sim$ 2.1 (Ciardullo et al. 2012, dot-dashed green), z $\sim$ 3.1 (Ciardullo et al. 2012, dotted blue), z $\approx$ 3.7 (Ouchi et al. 2008, magenta lines), with the contours denoting confidence levels of 68.3\% and 95.54\% ($\Delta \chi^2$ = 2.3 and 6.18, corresponding to 1$\sigma$ and 2$\sigma$), and fitting results of \lya\ LFs at z $\approx$ 5.7 (dark-cyan pentagon for Ouchi et al. 2010, cyan-blue pentagon for Kashikawa et al. 2011, and dark-magenta pentagon for Hu et al. 2010), and z $\approx$ 6.5 (black circle for Ouchi et al. 2010, cyan-blue circle for Kashikawa et al. 2011, and grey circle for Hu et al. 2010).  The abbreviation in the figure: C11: Ciardullo et al. (2012); W09: Wang et al. (2009); D07: Dawson et al. (2007); O08: Ouchi et al. (2008); K11: Kashikawa et al. (2011); H10: Hu et al. (2010); O10: Ouchi et al. (2010). Note that we fit all the best-fitting parameters with fixing $\alpha$ = -1.5 (except the z$\sim$5.7 and 6.5 data from Kashikawa et al. 2011, who did not give the differential \lya\ LFs and we quoted their fitting results),  so the L$^*$ and $\Phi^*$ discussed here are slightly different from those reported by Dawson et al. (2007, where $\alpha$ = -1.6) and Ciardullo et al. (2012, where  $\alpha$ = -1.65).}
\label{lyalffitz45}
\end{center}
\end{figure}

In Figure \ref{difflyalf} and \ref{lyalffitz45} we plot all the differential \lya\ luminosity functions and corresponding \lya\ luminosity function parameters L$^*$ and $\Phi^*$ at 2.1 $\leq$ z $\leq$ 6.6 from the literature (Ciardullo et al. 2012, Ouchi et al. 2008, Wang et al. 2009, Dawson et al. 2007, Ouchi et al. 2010, Kashikawa et al. 2011, Hu et al. 2010. For comparison, we also plot the differential \lya\ luminosity functions of GALEX selected LAE samples at z $\sim$ 0.3 and $\sim$ 1.0 from Cowie et al. 2010, 2011, which show significant differences to z $>$ 2 surveys.).  Though previous studies have implied that the \lya\ luminosity function did not evolve from 3 $< z <$ 6, Ciardullo et al. (2012) measured apparent evolution in L$^*$ and $\Phi^*$ from z = 3.1 to 2.1,  which are highlighted as the dashed contours in Figure \ref{lyalffitz45}. We find that from z=3.1 to 4.5, there is also apparent evolution (except in the ECDFS field). The best-fit luminosity function parameters at z $\sim$ 3.1 are located in the 2-$\sigma$ (95\%) confidence region of the parameters of the z $\sim$ 4.5 luminosity functions in ECDFS, while outside the 99\% confidence region of the parameters of two other z $\sim$ 4.5 fields. Thus it is difficult to judge the evolution of \lya\ luminosity function from z $\sim$ 3.1 to z $\sim$ 4.5. 
The luminosity function we report for $z=4.5$ is broadly similar to the mean of various reported luminosity functions at $z=5.7$ (Ouchi et al 2008, Hu et al 2010, Kashikawa et al 2011). Finally, between redshift z = 5.7 and 6.5,  several groups had reported evolution (Ouchi et al. 2010, Kashikawa et al. 2011, Hu et al. 2010) with fewer or fainter LAEs at z=6.5, but the parameters  of L$^*$ and $\Phi^*$ measured by different groups spanned a disturbingly large range.

\begin{table*}
\begin{minipage}{180mm}
\caption{\lya\ luminosity functions at z$\sim$ 4.5. Note that we fixed faint-end slope of $\alpha$ = -1.5 in the fitting.}
\begin{tabular}{@{}lccccccccc@{}}
\hline \hline
Fields 	&	Conf./Targ./Cand.	&	 Volume 	&	Depth(NB) & L$^*$ & $\Phi^*$ & Covar & L$^*\Phi^*$  &  $\chi^2$/dof & Ref.  \\
&   Number  &  Mpc$^3$  & AB mag & log$_{10}$(erg/s) & log$_{10}$(Mpc$^{-3}$) &  & log$_{10}$(erg/s Mpc$^{-3}$)  &\\
\hline
CDF-S 		&	 46 	/	 64		/	 112 		&	452,616 & 24.9 &42.75$\pm$0.09  &  -3.25$\pm$0.16 & $ \Bigl(\begin{array}{cc} 0.0086 & -0.014 \\ -0.014 & 0.025\end{array} \Bigr)$ & 39.50$\pm$0.07 & 6.5/6 & This work 	\\
Bootes		&	51 		/	 80		/	160		& 774,752  & 25.0 & 43.00$\pm0.11$ & -3.69$\pm0.13$ &$ \Bigl(\begin{array}{cc} 0.0114 & -0.0127 \\ -0.0127 & 0.0169\end{array} \Bigr)$ & 39.31$\pm$0.05 & 3.8/5 &  Dawson+07$^a$	 	\\
Cetus		 &	110 		/	 194		/	226 		& 617,622 & 24.8 & 42.72$\pm0.06$ & -3.38$\pm0.11$ & $ \Bigl(\begin{array}{cc} 0.0033 & -0.0057 \\ -0.0057 & 0.0117\end{array} \Bigr)$ &   39.34$\pm$0.06    & 9.4/7 &  Wang+09		\\
 \hline
Combine & 207 / 338 / 498 & 1,844,990 & 24.8 & 42.83$\pm$0.06 & -3.48$\pm$0.09 & $ \Bigl(\begin{array}{cc} 0.0038 & -0.0052 \\ -0.0052 & 0.0080\end{array} \Bigr)$ & 39.35$\pm$0.04  & 2.9/6 & This work \\
 \hline
\label{comblyalf}
\end{tabular}
\footnotetext[1]{Dawson et al. (2007) fitted the Schechter function by fixing faint-end-slope $\alpha$ = -1.6, here we fit their data with fixed faint-end-slope $\alpha$ = -1.5.}
\end{minipage}
\end{table*}

 \subsubsection{The Star Formation Rate Density of LAEs and the Global \lya\ Escape Fractions}
 \label{sec:lyalf:evolution}

With the above discussed \lya\ luminosity functions of LAEs at different redshifts, we can explore the star formation rate density (SFRD) from LAEs as a function of redshift. Assuming case B recombination (Brocklehurst 1971), the \lya\ luminosity density can be converted to a SFR density with a relationship of SFR [M$_\odot$ yr$^{-1}$] = 9.1$\times$10$^{-43}$ L(\lya) [ergs s$^{-1}$] (Kennicutt 1998).  The \lya\ luminosity density is the integration of the Schechter function (Equation 2) over luminosity, which gives:
\begin{eqnarray}
L_{(>L_{min})}  & = & \int_{L_{min}}^{\infty} L' \phi(L')dL',  \\ \nonumber
 & = & \Gamma(\alpha+2,L_{min}/L^*) \times L^*\Phi^* ,
\end{eqnarray}
here $\alpha$ is the faint-end slope, and $\Gamma(\alpha+2,L_{min}/L^*)$ is the incomplete gamma function. So $\Gamma(\alpha+2,L_{min}/L^*)$ $\times$ L$_*$$\Phi^*$ should be an independent parameter and it presents the density of \lya\ luminosities at a specific redshift with a fixed ratio of $L_{min}/L^*$ and a fixed faint-end slope of $\alpha$. When integrating from zero to infinity, $\Gamma(\alpha+2,L_{min}/L^*)$ = [1.30, 1.77, 2.55, 4.59] for $\alpha$ = [-1.3, -1.5, -1.65, -1.8], and integrating from $L_{min}/L^*$ = 0.1 to infinity, $\Gamma(\alpha+2,L_{min}/L^*)$ = [1.02, 1.16, 1.30, 1.49] for $\alpha$ = [-1.3, -1.5, -1.65, -1.8].  In the following discussion, we fix $\alpha$ = -1.5 and $L_{min}/L^*$ = 0 (i.e., we integrate the luminosity function down to zero luminosity). The parameter L$_*$$\Phi^*$ is discussed in previous section (Section 4.3.1), and due to the covariance of L$_*$ and $\Phi^*$, the uncertainty on L$_*$$\Phi^*$ is significantly lower than the individual error on L$_*$ or $\Phi^*$. However, previous works did not give the covariance matrix in the luminosity function fitting, which is necessary for the error-estimate of L$_*$$\Phi^*$. We thus re-fit the \lya\ luminosity functions of LAE surveys from the literature (using published number densities) with a fixed faint-end slope of $\alpha$ = -1.5 to get the values and errors of L$_*$, $\Phi^*$, L$_*$$\Phi^*$, as well as the covariance matrixes of L$_*$ and $\Phi^*$ (see Table \ref{allnblyalf}). 
 
In Figure \ref{zlphi} we plot the SFRDs (\lya\ luminosity densities of $\Gamma(0.5, 0)$ $\times$ L$_*$(\lya)$\Phi^*$) for different narrowband surveys at corresponding redshifts. For a comparison, we also plot the observed and dust-corrected SFRDs from UV luminosity densities of LBGs (Bouwens et al. 2012). The integrated \lya\ density for GALEX selected LAEs at z $\sim$ 0.3 and $\sim$0.95 by Cowie et al. (2010, 2011) are also added. The \lya\ luminosity density for LAEs shows a qualitatively similar evolution to the global SFR density at 0 $<$ z $\leq$ 4.5 (see Figure \ref{zlphi} ), and also at z = 5.7 and 6.5, albeit with large uncertainties. If we use the results from Hu et al. for LAEs at z = 5.7 and 6.5 only, the \lya\ luminosity density evolves similar to the global dust-corrected SFR density in the whole redshift range 0 $<$ z $\leq$ 6.5, in that it continues declining out to the highest redshifts.  If instead we consider the \lya\ luminosity function results at z = 5.7 and 6.5 from Ouchi et al. (2010) and Kashikawa et al. (2011), LAEs seem to take an increasingly dominant role as z $\rightarrow$ 6.5.  Recent spectroscopic observations of galaxies at 3 $<$ z $<$ 7 have found that the fraction of galaxies exhibiting \lya\ in emission is increasing with redshift from 4 $< z <$ 6 (Stark et al. 2011).  This observation is consistent with SFR density plot considering the results from Ouchi et al. (2010) and Kashikawa et al. (2011), as their SFR densities comprise very high percentages of the dust-corrected SFR density at z = 5.7 and 6.5.  If this is truly the case, it appears as if the \lya\ luminosity density is roughly constant at $z >$ 3.  However, as the high-redshift end is quite uncertain, more surveys at z $\geq$ 6.5 are needed to confirm this result.    


The ratio of the SFRDs from the integrated \lya\ luminosity density of LAEs to those from the dust-corrected UV SFRDs can be defined as the global \lya\ escape fraction, which is the fraction of \lya\ photons which escape galaxies divided by the number created in the star-forming regions. Unlike the \lya\ escape fraction estimated in Section 4.1, which is the  \lya\ escape fraction for LAE galaxies only, the global \lya\ escape fraction tells the same quantity, but averaged over all star-forming galaxies (hence, the number by definition will be lower). We plot the global \lya\ escape fraction as a function of redshift in Figure \ref{zsfrd}, with the comparison using UV luminosity density integration limits of 0.3L$^*_{z=3}$ (Bouwens et al. 2007) or 0.05L$^*_{z=3}$ (Bouwens et al. 2012) shown separately. Hayes et al. (2011) found an evolved global \lya\ escape fraction as a function of redshift, which is shown in a power-law form of $f^{Ly\alpha}_{esc}$ $\propto$ (1+z)$^{2.6\pm0.2}$ at 0 $<$ z $<$ 6 (see green solid line in Figure \ref{zsfrd}). 

Note that since the analysis in Hayes et al., a number of new luminosity function results have been published: the luminosity functions of GALEX selected LAEs at z $\sim$ 0.3 and 1.0 were updated by Cowie et al. (2010, 2011) and Barger et al. (2012), \lya\ luminosity functions at z = 2.1 and 3.1 were updated by Ciardullo et al. (2012), new luminosity functions at z$\sim$ 4.5 are presented in this work, and LAE surveys at z = 5.7 and 6.5 were performed by Hu et al. (2010), Ouchi et al. (2010) and Kashikawa et al. (2011).  Remarkably, the global \lya\ escape fraction relation from Hayes et al. is seemingly consistent with all new results, assuming a dust-corrected UV luminosity density integrated down to 0.05L$^*_{LBG, z=3}$,  with the exception of those from Hu et al. (2010), though only at the $\sim$ few sigma level.
The results from Hu et al. imply that the \lya\ escape fraction should not evolve much from 4 $< z <$ 6.5, while the Ouchi et al. (2010) and Kashikawa et al. (2011) results support a significant increase of the fraction to z = 6.5.



These results seemingly support a scenario where as one goes to higher redshift, more star-forming galaxies would be classified as LAEs.  Hayes et al. (2011) connected the increase of global \lya\ escape fraction as a function of redshift to a decrease in the typical dust attenuation. SED fitting and spectroscopic confirmation of high-z LBGs support this (e.g., Stark et al. 2011; Finkelstein et al.\ 2012; Bouwens et al.\ 2012).  However, the results of Hu et al. (2010) at z = 5.7 and 6.5 are not consistent with this scenario.  We note that at $z >$ 6, the existence of neutral hydrogen in the IGM may affect the observation of \lya\, and cosmic variance, as shown in our z$\sim$4.5 samples where the IGM should not be a factor, may also have a significant effect on the difference of \lya\ luminosity functions at z $\geq$ 6.  Thus more samples at z $\sim$ 5.7, 6.5 and $\gtrsim$7 (though until now there are only a few spectroscopic confirmed galaxies at z$\geq$ 7; Iye et al. 2006, Rhoads et al. 2012) are needed to robustly examine the evolution of the global \lya\ escape fraction at the highest redshifts.


\begin{figure*}
\begin{center}
\includegraphics[totalheight=0.65\textheight]{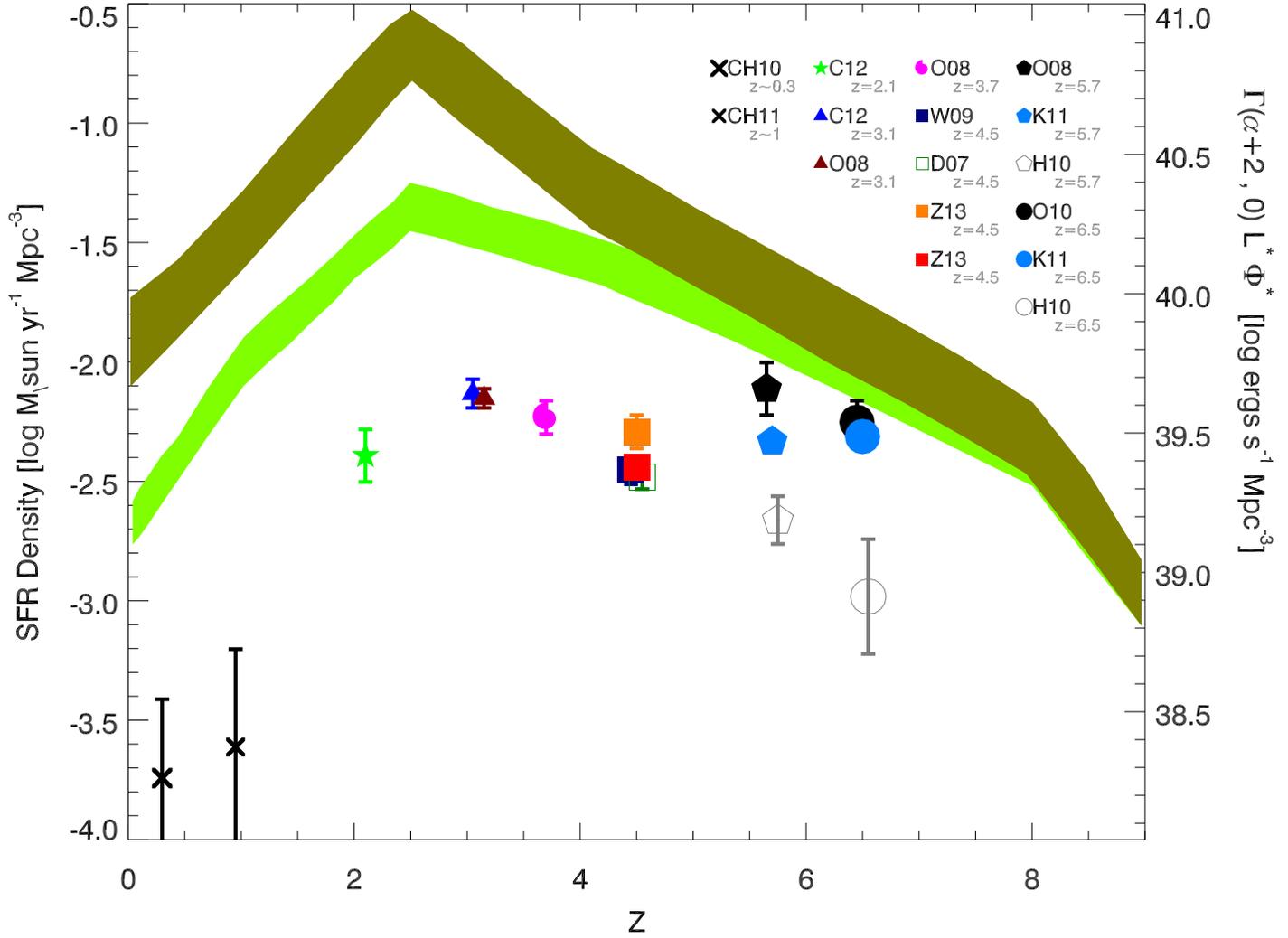} 
\caption{The evolution of $\Gamma(\alpha+2,0)$L$_*$$\Phi^*$ for LAEs, and a comparison to to the UV SFR density of the universe integrated down to 0.3 L$^*_{LBG, z=3}$ with or without dust correction (dark yellow or dark cyan filled regions) from Bouwens et al. (2012). We use the Kennicutt (1998) relation and case B recombination to convert from L$^*$(\lya) $\times$ $\Phi^*$ to SFR density. All the points at z$>$ 2 are from ground-based narrowband surveys, and points at z$\sim$ 0.3 and 0.95 are from GALEX selected LAEs by Cowie et al. (2010, 2011).  The symbols are same to the previous figure,  CH10: Cowie et al. 2010; CH11: Cowie et al. 2011; C11: Ciardullo et al. (2012);W09: Wang et al. (2009); D07: Dawson et al. (2007); O08: Ouchi et al. (2008); S06: K11: Kashikawa et al. (2011); H10: Hu et al. (2010); O10: Ouchi et al. (2010). Note that all the best-fitting parameters are obtained by fixing $\alpha$ = -1.5.   }
\label{zlphi}
\end{center}
\end{figure*}

\begin{figure*}
\begin{center}
\includegraphics[totalheight=0.65\textheight]{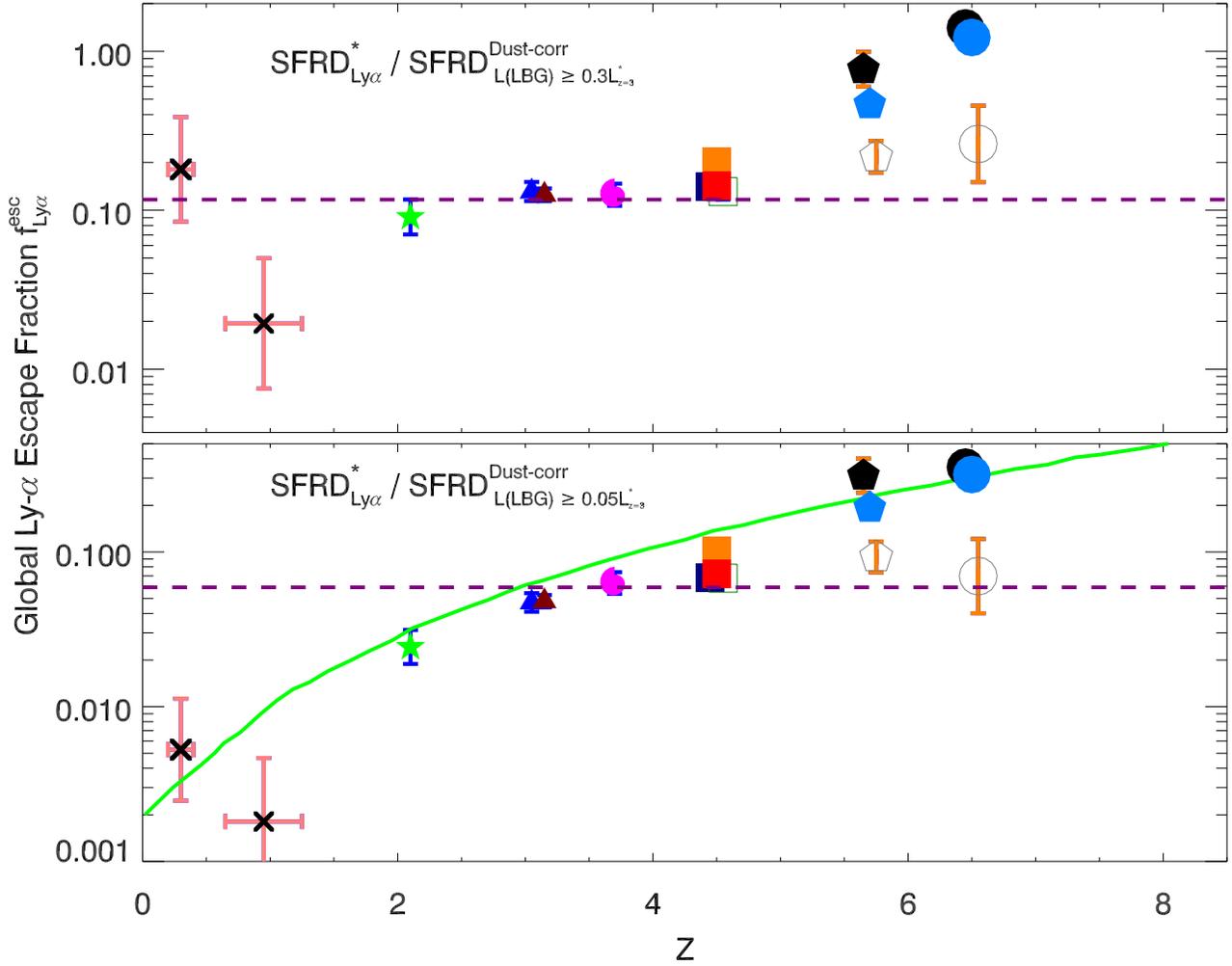} 
\caption{The ratio of the SFR densities from LAE surveys to the dust-corrected UV SFR densities (the global \lya\ escape fraction) as a function of redshift. The symbols are same as in Figure \ref{zlphi}. The dust-corrected SFR densities are converted  from the UV luminosity densities of LBGs integrated down to 0.3L$^*_{z=3}$ (top, Bouwens et al. 2007) or 0.05L$^*_{z=3}$ (bottom, Bouwens et al. 2012), and should correspond to the total SFR densities. The errors in the ratio is only from the SFR density of LAE surveys. The y-axis can be presented as the global \lya\ escape fraction, and the average values (dashed and dotted purple lines) at 2 $< $ z $<$ 5 are 11.7\% and 5.9\% when the total dust-corrected SFR density integrated to 0.3L$^*_{z=3}$ and 0.04L$^*_{z=3}$, respectively. The green solid line is the global \lya\ escape fraction from Hayes et al. (2011). }
\label{zsfrd}
\end{center}
\end{figure*}
    
\begin{table*}
\begin{minipage}{220mm}
\caption{\lya\ luminosity functions at different redshifts. Note that we fixed faint-end slope of $\alpha$ = -1.5 in all the fittings below except the \lya\ LFs of Kashikawa et al. 2011, whose parameters were directly from their paper.}
\begin{tabular}{@{}llcccccc@{}}
\hline \hline
Redshift & Fields 	 & L$^*$ & $\Phi^*$ & Covar & L$^*\Phi^*$  &  $\chi^2$/dof & Ref.  \\
& &    log$_{10}$(erg/s) & log$_{10}$(Mpc$^{-3}$) &  & log$_{10}$(erg/s Mpc$^{-3}$)  &\\ \hline
$\sim$0.3 & GALEX    & 41.53$\pm$0.16 & -3.48$\pm$0.48 &  $ \Bigl(\begin{array}{cc} 0.026 & -0.075 \\
-0.075 & 0.229  \end{array} \Bigr)$ & 38.05$\pm$0.33 & 1.7/4 & Cowie+10 \\
$\sim$1 & GALEX       & 42.56$\pm$0.16 & -4.38$\pm$0.55 & $ \Bigl(\begin{array}{cc} 0.027 & -0.083 \\
-0.083 & 0.304 \end{array} \Bigr)$ & 38.18$\pm$0.41   & 0.0/0  &  Cowie+11 \\
\hline
2.1 & CDF-S 	       & 42.18$\pm$0.12 & -2.78$\pm$0.23 & $ \Bigl(\begin{array}{cc} 0.0151 & -0.0270 \\ -0.0270 & 0.0511\end{array} \Bigr)$ & 39.40$\pm$0.11 & 5.4/6 & Ciardullo+12 \\
3.1 & CDF-S 	   & 42.62$\pm$0.09 & -2.96$\pm$0.14 & $ \Bigl(\begin{array}{cc} 0.0088 & -0.0126 \\ -0.0126 & 0.0200\end{array} \Bigr)$ & 39.66$\pm$0.06 & 4.2/9 & Ciardullo+12 \\
3.1 & SXDS        & 42.80$\pm$0.06 & -3.17$\pm$0.08 & $ \Bigl(\begin{array}{cc} 0.0033 & -0.0045 \\ -0.0045 & 0.0071\end{array} \Bigr)$ & 39.64$\pm$0.04 & 5.3/6 & Ouchi+08 \\
3.7 & SXDS           & 42.86$\pm$0.07 & -3.30$\pm$0.13 & $ \Bigl(\begin{array}{cc} 0.0052 & -0.0087 \\ -0.0087 & 0.0166\end{array} \Bigr)$ &  39.56$\pm$0.07 & 1.3/3 & Ouchi+08 \\ \hline
4.5 & CDF-S 		& 42.75$\pm$0.09  &  -3.25$\pm$0.16 & $ \Bigl(\begin{array}{cc} 0.009 & -0.014 \\ -0.014 & 0.025\end{array} \Bigr)$ & 39.50$\pm$0.07 & 6.5/6 & This work 	\\
4.5 & Bootes		&43.00$\pm0.11$ & -3.69$\pm0.13$ &$ \Bigl(\begin{array}{cc} 0.0114 & -0.0127 \\ -0.0127 & 0.0169\end{array} \Bigr)$ & 39.31$\pm$0.05 & 3.8/5 &  Dawson+07$^a$	 	\\
4.5 & Cetus		 &42.72$\pm0.06$ & -3.38$\pm0.11$ & $ \Bigl(\begin{array}{cc} 0.0033 & -0.0057 \\ -0.0057 & 0.0117\end{array} \Bigr)$ &   39.34$\pm$0.06    & 9.4/7 &  Wang+09		\\
4.5 & Combine & 42.83$\pm$0.06 & -3.48$\pm$0.09 & $ \Bigl(\begin{array}{cc} 0.0038 & -0.0052 \\ -0.0052 & 0.0080\end{array} \Bigr)$ & 39.35$\pm$0.04  & 2.9/6 & This work \\ \hline
%
%
%
5.7 & SXDS      & 42.86$\pm$0.10 & -3.17$\pm$0.19 &  $ \Bigl(\begin{array}{cc} 0.0100 & -0.0180 \\ -0.0180 & 0.0378\end{array} \Bigr)$ &  39.68$\pm$0.11 & 2.8/4 & Ouchi+08 \\
5.7 &     SDF        & 43.02$\pm$0.06 & -3.56$\pm$0.09 & $ \Bigl(\begin{array}{cc} 0.0036 & -0.0057 \\ -0.0057 & 0.0081 \end{array} \Bigr)$ &  39.46$\pm$0.02 & -- & Kashikawa+11 \\
5.7 &   multiple      & 42.93$\pm$0.13 & -3.80$\pm$0.22 & $ \Bigl(\begin{array}{cc} 0.0175 & -0.0276 \\ -0.0276 & 0.0469\end{array} \Bigr)$ &  39.13$\pm$0.10 & 0.135/2 & Hu+10 \\
 6.5 & SXDS        & 42.70$\pm$0.10 & -3.16$\pm$0.18 &  $ \Bigl(\begin{array}{cc} 0.0108 & -0.0183 \\ -0.0183 & 0.0339\end{array} \Bigr)$ &  39.54$\pm$0.09 & 1.3/3 & Ouchi+10\\ 
 6.5 &      SDF      & 42.76$\pm$0.10 & -3.28$\pm$0.20 & $ \Bigl(\begin{array}{cc} 0.010 & -0.024 \\ -0.024 & 0.040 \end{array} \Bigr)$ &  39.48$\pm$0.02 & -- & Kashikawa+11 \\
6.5 &   multiple        & 42.99$\pm$0.23 & -4.18$\pm$0.45 &  $ \Bigl(\begin{array}{cc} 0.0533 & -0.0986 \\ -0.0986 & 0.2028\end{array} \Bigr)$ &  38.81$\pm$0.24 & 0.004/1 & Hu+10 \\
\hline
\label{allnblyalf}
\end{tabular}
\end{minipage}
\end{table*}

\section{CONCLUSION}
\label{sec:conclusion}

We present a sample of spectroscopically confirmed LAEs at z $\sim$ 4.5 in the ECDFS field. This sample is from two contiguous narrowband images (NB665 and NB673), and a much smaller region of a shallower narrowband image (NB656). The main scientific results are summarized as below:

\begin{itemize}
\item We identify 46  z $\sim$ 4.5 LAEs, 5 LBGs and 1 [O\,{\sc iii}] emitter from our targeted 64 LAE candidates. The \lya\ confirmation fraction is $\sim$70\%-80\%, and the contamination fraction is $\sim$14\%-19\%. All targets with f$_{Ly\alpha}$ $>$ 3.7$\times$10$^{-17}$ erg cm$^2$s$^{-1}$ are confirmed. 

\item We do not find any C\,{\sc iv} or He\,{\sc ii} lines  in the spectra of the confirmed LAEs, and after stacking 29 spectra we obtain 2-$\sigma$ upper limits of $f(\mbox{C\,{\sc iv}}\lambda 1549)/f(\mbox{Ly-}\alpha ) <$ 3.7\% and $f(\mbox{He\,{\sc ii}}\lambda 1640)/f(\mbox{Ly-}\alpha ) <$ 3.4\%.
. 
\item The quasar previously confirmed as a type 1 AGN at z = 4.48 in our sample is also detected with deep X-ray and radio observations. The remaining LAEs are not detected in X-ray or radio even with stacking methods, and the stacked upper limits in radio and X-ray can be converted to average SFR as SFR$_{Radio}$ $<$ 100 M$_\odot$ and SFR$_{X}$ $<$ 214 M$_\odot$ at a 95.4\% confidence level. 

\item The upper limit of the He\,{\sc ii} to \lya\ line ratio in our coadded spectrum can be converted to a 2-$\sigma$ upper limit on SFR$_{PopIII}$ $<$ 0.3 M$_\odot$ yr$^{-1}$. 
This is only $<$ 0.3\% (1.25\%) of the total (observed) SFR at z$\sim$ 4.5, implying that Pop III stars at z $\sim$ 4.5 are very rare. 

\item Our \lya\ luminosity functions are consistent with previous surveys at 3 $\leq z \leq$ 6.5, as log$_{10}$(L$^{*}$) =  42.75 $\pm$ 0.09 and log$_{10}$($\Phi^*$) = -3.25 $\pm$ 0.16. However, our \lya\ luminosity functions in the two narrowband images differs at $>$ 90\% confidence level, at least partially due to cosmic variance. 

\item The combined z$\sim$ 4.5 \lya\ luminosity functions have log$_{10}$(L$^{*}$) =  42.83 $\pm$ 0.06 and log$_{10}$($\Phi^*$) = -3.48 $\pm$ 0.09. The SFRD of LAEs estimated from integrating the \lya\ luminosity functions over luminosity for different narrowband surveys shows a trend that differs from that of the UV-derived SFRD at $z >$ 6, in that LAEs appear to be taking on an increasingly dominant role.  This may be due to a decrease in dust attenuation, allowing more \lya\ photons to escape a given galaxy.

\end{itemize}

\section*{Acknowledgments}
We gratefully thank two anonymous referees for the insightful suggestions and comments that helped us to improve the paper significantly. We would like to thank the support of NSF grant AST-0808165 and NOAO TSIP program.  The work of JXW is supported by National Basic Research Program of China (973 program, Grant No. 2007CB815404),  and Chinese National Science Foundation (Grant No. 10825312, 11233002).

\clearpage

\begin{table*}
\vskip-9pt
\hskip -4cm
\small
\begin{minipage}{100mm}
\caption{Spectroscopically Confirmed \lya\ Emitters at z$\sim$ 4.5 in the CDF-S field}
\begin{tabular}{@{} lccccccccccc@{}}
\hline \hline
Obj & 	Ra & 	Dec & 	Redshift & Mask &	F$_{Ly\alpha}^1$   & EW$_{rest}^2$ & Spec.$^3$ & FWHM$^4$ & a$_{\lambda}$ & a$_{flux}$\\
 &  &   &   &   &   & (\AA ) &  grade & (\AA) &  & \\ 
 \hline
CH8-1--665-24 & 53.20420 & -27.81722 & 4.431 & 4 & 2.66$\pm$0.54 &  149.1$^{+  36.4}_{-  36.4}$ & 1 &   11.0$^{+   0.7}_{-   0.8}$ &    0.9$^{+   0.0}_{-   0.0}$ &    0.7$^{+   0.0}_{-   0.0}$  \\ CH8-2--665-26 & 53.22515 & -27.83356 & 4.431 & 3 & 1.98$\pm$0.65 &   56.3$^{+  11.0}_{-  11.0}$ & 1 &   12.3$^{+   3.3}_{-   2.8}$ &    1.3$^{+   0.8}_{-   0.5}$ &    1.4$^{+   0.2}_{-   0.2}$  \\ CHa-2--656-2 & 53.16571 & -27.85415 & 4.411 & 3 & 3.93$\pm$0.40 & $>$10000 & 1 &   11.1$^{+   0.7}_{-   0.4}$ &    2.2$^{+   0.5}_{-   0.7}$ &    0.9$^{+   0.1}_{-   0.1}$  \\ CHa-3--656-3 & 53.24325 & -27.89433 & 4.391 & 3 & 4.15$\pm$0.42 & $>$10000 & 1 &   10.1$^{+   0.6}_{-   0.9}$ &    1.5$^{+   0.8}_{-   0.4}$ &    1.3$^{+   0.2}_{-   0.2}$  \\ CHa-4--656-4 & 53.20102 & -27.86025 & 4.364 & 3 & 3.11$\pm$0.34 &  124.6$^{+  29.5}_{-  27.5}$ & 2 &    8.3$^{+   8.8}_{-   1.3}$ &    3.0$^{+   3.0}_{-   1.0}$ &    3.3$^{+   1.0}_{-   0.8}$  \\ CS2-1--673-35 & 53.03973 & -27.77323 & 4.512 & 3 & 2.42$\pm$0.63 & $>$10000 & 1 &    5.4$^{+   4.4}_{-   1.8}$ &    1.1$^{+   1.5}_{-   0.9}$ &    0.9$^{+   1.2}_{-   0.7}$  \\ 
CS2-2--673-36 & 53.06737 & -27.81233 & 4.542 & 3 & 3.46$\pm$0.62 &  189.7$^{+  46.8}_{-  46.8}$ & 1 &   10.0$^{+   0.3}_{-   0.4}$ &    1.0$^{+   0.3}_{-   0.4}$ &    0.7$^{+   0.1}_{-   0.1}$  \\ 
CS2-3--673-39 & 53.10224 & -27.79322 & 4.500 & 3 & 2.47$\pm$0.47 &  166.6$^{+  42.9}_{-  42.9}$ & 1 &    9.4$^{+   0.8}_{-   0.9}$ &    1.5$^{+   0.2}_{-   0.4}$ &    1.5$^{+   0.1}_{-   0.1}$  \\ 
CS2-4--673-42 & 53.11923 & -27.93295 & 4.527 & 3 & 2.27$\pm$0.37 &  111.4$^{+  11.3}_{-  11.7}$ & 1 &    7.4$^{+   2.4}_{-   0.7}$ &    2.7$^{+   1.8}_{-   0.9}$ &    1.3$^{+   0.3}_{-   0.3}$  \\ 
CS2-5--673-44 & 53.13542 & -27.72973 & 4.500 & 3 & 2.57$\pm$0.37 &   42.4$^{+   5.3}_{-   5.5}$ & 1 &    6.4$^{+   4.4}_{-   2.6}$ &    0.9$^{+   0.7}_{-   0.5}$ &    1.0$^{+   1.1}_{-   0.6}$  \\ 
CS2-8--673-52 & 53.16770 & -27.88614 & 4.500 & 4 & 2.05$\pm$0.36 &  703.0$^{+ 497.0}_{- 262.0}$ & 2 &   11.2$^{+   1.5}_{-   5.9}$ &    1.2$^{+   1.3}_{-   0.7}$ &    1.3$^{+   1.2}_{-   0.5}$  \\ 
673-5--665-11 & 52.90865 & -27.86655 & 4.505 & 5 & 5.80$\pm$0.58 & $>$10000 & 1 &    9.3$^{+   1.6}_{-   0.7}$ &    1.1$^{+   0.0}_{-   0.0}$ &    1.3$^{+   0.1}_{-   0.1}$  \\ 
673-16--665-13 & 52.94718 & -27.89357 & 4.507 & 4 & 5.87$\pm$0.59 &   37.6$^{+   9.0}_{-   9.0}$ & 1 &   10.1$^{+   1.2}_{-   3.1}$ &    1.6$^{+   0.4}_{-   0.5}$ &    0.9$^{+   0.1}_{-   0.1}$  \\ 
665-31 & 53.29958 & -27.91413 & 4.458 & 5 & 2.86$\pm$0.86 & $>$10000 & 1 &   15.9$^{+   0.8}_{-   1.7}$ &    2.5$^{+   2.5}_{-   0.8}$ &    3.0$^{+   1.0}_{-   0.7}$  \\ 
665-34 & 53.32557 & -27.79347 & 4.480 & 3 & 3.28$\pm$0.71 &   48.1$^{+  13.7}_{-  13.7}$ & 1 &   10.3$^{+   0.7}_{-   0.4}$ &    0.8$^{+   1.1}_{-   0.0}$ &    1.3$^{+   0.2}_{-   0.2}$  \\ 
665-39 & 53.35505 & -27.81548 & 4.456 & 3 & 5.33$\pm$0.72 & $>$10000 & 1 &   10.0$^{+   0.5}_{-   0.8}$ &    1.2$^{+   0.0}_{-   0.3}$ &    1.2$^{+   0.1}_{-   0.1}$  \\ 
673-3--665-8 & 52.89650 & -27.86801 & 4.505 & 5 & 4.73$\pm$0.58 &   85.2$^{+  10.3}_{-  10.3}$ & 1 &   10.3$^{+   0.4}_{-   0.4}$ &    1.0$^{+   0.3}_{-   0.2}$ &    0.7$^{+   0.0}_{-   0.0}$  \\ 
673-10 & 52.92503 & -27.81979 & 4.515 & 2 & 2.28$\pm$0.33 & $>$10000 & 2 &    9.0$^{+   0.2}_{-   0.2}$ &    2.5$^{+   0.0}_{-   1.2}$ &    1.5$^{+   0.2}_{-   0.3}$  \\ 
673-11 & 52.92717 & -27.76996 & 4.520 & 3 & 3.11$\pm$0.62 &   36.4$^{+   8.9}_{-   9.1}$ & 1 &   12.4$^{+   4.1}_{-   3.5}$ &    2.0$^{+   0.0}_{-   0.3}$ &    1.0$^{+   0.1}_{-   0.1}$  \\ 
673-13 & 52.93424 & -27.78658 & 4.520 & 5 & 4.17$\pm$0.38 &   21.8$^{+   5.5}_{-   5.4}$ & 1 &   12.2$^{+   2.9}_{-   0.9}$ &    1.7$^{+   0.0}_{-   0.4}$ &    0.9$^{+   0.1}_{-   0.1}$  \\ 
673-24 & 52.97977 & -27.71705 & 4.532 & 3 & 2.02$\pm$0.41 &  206.0$^{+  19.2}_{-  18.9}$ & 2 &    7.3$^{+   2.4}_{-   2.7}$ &    1.3$^{+   0.5}_{-   0.5}$ &    1.0$^{+   0.3}_{-   0.3}$  \\ 
673-28 & 53.00088 & -27.82576 & 4.512 & 4 & 1.98$\pm$0.34 &   65.8$^{+  18.8}_{-  18.3}$ & 1 &    8.8$^{+   1.8}_{-   0.5}$ &    0.9$^{+   0.7}_{-   0.4}$ &    1.1$^{+   0.3}_{-   0.3}$  \\ 
673-37 & 53.08812 & -27.99047 & 4.525 & 3 & 1.97$\pm$0.62 & $>$10000 & 1 &    9.6$^{+   2.0}_{-   1.1}$ &    1.1$^{+   0.4}_{-   0.0}$ &    1.4$^{+   0.2}_{-   0.2}$  \\ 
673-60 & 53.22586 & -27.98279 & 4.500 & 4 & 2.98$\pm$0.64 &  728.4$^{+ 317.8}_{- 210.7}$ & 1 &   14.4$^{+   1.5}_{-   1.3}$ &    1.0$^{+   0.3}_{-   0.0}$ &    1.1$^{+   0.2}_{-   0.2}$  \\ 
673-61 & 53.23241 & -27.96140 & 4.517 & 5 & 3.71$\pm$0.64 & $>$10000 & 1 &    8.3$^{+   1.0}_{-   0.7}$ &    3.2$^{+   1.2}_{-   1.8}$ &    1.6$^{+   0.5}_{-   0.4}$  \\ 
673-66 & 53.33446 & -27.94221 & 4.515 & 4 & 2.54$\pm$0.65 &   35.0$^{+   4.5}_{-   4.7}$ & 1 &    9.9$^{+   1.1}_{-   1.4}$ &    2.1$^{+   0.5}_{-   0.7}$ &    1.6$^{+   0.4}_{-   0.3}$  \\ 
673-69 & 53.34030 & -27.94488 & 4.515 & 3 & 2.95$\pm$0.63 &   33.5$^{+   7.8}_{-   7.8}$ & 1 &   10.3$^{+   3.7}_{-   4.9}$ &    0.9$^{+   0.3}_{-   0.2}$ &    1.1$^{+   0.3}_{-   0.2}$  \\ 
673-74 & 53.35577 & -27.79243 & 4.525 & 4 & 3.56$\pm$0.36 &   27.0$^{+   1.2}_{-   1.3}$ & 1 &    8.5$^{+   0.9}_{-   0.2}$ &    0.9$^{+   1.0}_{-   0.4}$ &    1.0$^{+   0.4}_{-   0.3}$  \\ 
673-7 & 52.91647 & -27.87911 & 4.500 & 4 & 2.75$\pm$0.33 & 3130.6$^{+6012.2}_{-2173.5}$ & 1 &   11.5$^{+   1.9}_{-   2.8}$ &    1.2$^{+   0.0}_{-   0.0}$ &    0.9$^{+   0.2}_{-   0.1}$  \\ 
CS2-7-673-51 & 53.16261 & -27.80360 & 4.528 & 2 & 2.56$\pm$0.40 &  237.3$^{+ 112.1}_{-  82.4}$ & 1 &   12.4$^{+   0.7}_{-   0.9}$ &    1.4$^{+   0.0}_{-   0.0}$ &    1.2$^{+   0.1}_{-   0.1}$  \\ 
665-10 & 52.90203 & -27.71201 & 4.475 & 2 & 2.87$\pm$0.42 & $>$10000 & 2 &    7.5$^{+   4.0}_{-   0.4}$ &    1.4$^{+   0.0}_{-   0.7}$ &    1.0$^{+   0.1}_{-   0.2}$  \\ 
665-14 & 52.99119 & -28.00636 & 4.484 & 2 & 3.09$\pm$0.58 & $>$10000 & 1 &    9.0$^{+   0.3}_{-   2.2}$ &    0.6$^{+   0.0}_{-   0.0}$ &    1.2$^{+   0.1}_{-   0.1}$  \\ 
665-9 & 52.90171 & -27.75258 & 4.454 & 2 & 5.42$\pm$0.68 &   20.4$^{+   5.2}_{-   5.4}$ & 1 &   10.9$^{+   0.1}_{-   0.1}$ &    1.3$^{+   0.0}_{-   0.4}$ &    1.2$^{+   0.0}_{-   0.1}$  \\ 
673-12 & 52.93412 & -27.99292 & 4.500 & 2 & 2.50$\pm$0.50 & $>$10000 & 2 &   10.3$^{+   0.2}_{-   2.4}$ &    0.8$^{+   0.0}_{-   0.2}$ &    1.1$^{+   0.1}_{-   0.1}$  \\ 
673-14 & 52.93851 & -27.69510 & 4.505 & 2 & 2.22$\pm$0.61 & $>$10000 & 3 &    8.6$^{+   0.2}_{-   0.2}$ &    0.9$^{+   0.0}_{-   0.2}$ &    1.1$^{+   0.2}_{-   0.1}$  \\ 
673-17 & 52.94978 & -27.81017 & 4.516 & 2 & 6.48$\pm$0.61 &   69.3$^{+  15.8}_{-  15.5}$ & 1 &   10.6$^{+   2.0}_{-   0.4}$ &    1.6$^{+   0.0}_{-   0.0}$ &    1.0$^{+   0.0}_{-   0.0}$  \\ 
673-21 & 52.96696 & -27.72211 & 4.518 & 2 & 7.17$\pm$0.61 &   85.2$^{+   9.8}_{-  10.2}$ & 1 &   10.0$^{+   1.3}_{-   1.2}$ &    1.5$^{+   0.0}_{-   0.2}$ &    1.1$^{+   0.1}_{-   0.1}$  \\ 
673-22 & 52.97540 & -27.69971 & 4.493 & 2 & 2.07$\pm$0.35 &  109.2$^{+  34.2}_{-  31.9}$ & 1 &    8.4$^{+   0.3}_{-   0.4}$ &    0.9$^{+   0.3}_{-   0.4}$ &    0.8$^{+   0.1}_{-   0.0}$  \\ 
673-4 & 52.89806 & -27.67007 & 4.521 & 2 & 2.09$\pm$0.63 &   86.7$^{+  22.9}_{-  21.8}$ & 3 &    5.4$^{+   0.3}_{-   0.2}$ &    0.7$^{+   0.4}_{-   0.0}$ &    1.0$^{+   0.2}_{-   0.2}$  \\ 
CS2-6-673-46 & 53.13893 & -27.69562 & 4.528 & 2 & 1.93$\pm$0.61 &  134.1$^{+  23.1}_{-  22.5}$ & 1 &    8.6$^{+   0.8}_{-   1.2}$ &    1.7$^{+   0.0}_{-   0.6}$ &    0.9$^{+   0.1}_{-   0.1}$  \\ 
673-8 & 52.91662 & -27.75173 & 4.508 & 2 & 4.05$\pm$0.63 & $>$10000 & 1 &   10.3$^{+   1.9}_{-   0.6}$ &    0.7$^{+   0.2}_{-   0.0}$ &    1.1$^{+   0.0}_{-   0.0}$  \\ 
665-27 & 53.24300 & -27.56236 & 4.452 & 1 & 3.34$\pm$0.45 & $>$10000 & 3 &    7.2$^{+   0.4}_{-   0.4}$ &    1.2$^{+   0.3}_{-   0.5}$ &    0.8$^{+   0.1}_{-   0.1}$  \\ 
665-30 & 53.28141 & -27.56315 & 4.480 & 1 & 6.09$\pm$0.71 &  157.0$^{+  55.8}_{-  48.2}$ & 2 &   12.2$^{+   0.2}_{-   3.9}$ &    0.6$^{+   0.9}_{-   0.0}$ &    1.0$^{+   0.7}_{-   0.1}$  \\ 
665-40 & 53.36504 & -27.75002 & 4.482 & 1 & 4.20$\pm$0.66 &  132.7$^{+  17.7}_{-  16.3}$ & 1 &   10.5$^{+   0.3}_{-   0.3}$ &    1.0$^{+   1.0}_{-   0.0}$ &    1.3$^{+   0.2}_{-   0.2}$  \\ 
673-54 & 53.17277 & -27.64484 & 4.549 & 1 & 3.90$\pm$0.64 & $>$10000 & 1 &   10.1$^{+   0.9}_{-   0.8}$ &    1.5$^{+   0.0}_{-   0.0}$ &    1.3$^{+   0.1}_{-   0.1}$  \\ 
673-72 & 53.34993 & -27.83262 & 4.516 & 1 & 4.48$\pm$0.59 &  286.2$^{+ 148.3}_{- 105.2}$ & 1 &   12.2$^{+   0.4}_{-   2.3}$ &    0.7$^{+   0.1}_{-   0.3}$ &    0.7$^{+   0.1}_{-   0.3}$  \\ 
\hline
\label{data}
\end{tabular}
\footnotetext[1]{ The \lya\ line flux is calculated from the narrowband and broadband photometry in the unit of $10^{-17}$ erg s$^{-1}$ cm$^{-2}$.}
\footnotetext[2]{ LAEs with non-detection in the R-band are marked with  $EW > 10000$\AA. } 
\footnotetext[3]{LAEs confirmed from their spectra are assigned grades, here 1 means very good quality, 2 means not very significant, and 3 means low s/n. } 
\footnotetext[4]{The \lya\ line widths are directly measured from their spectra and not corrected for the instrument profiles. } 
\end{minipage}
\end{table*}

\clearpage

\bibliography{lyalfv8}


\label{lastpage}

\end{document}